\documentclass[graybox]{svmult}

\usepackage{url}
\usepackage[strings]{underscore} 
\usepackage{mathptmx}       
\usepackage{helvet}         
\usepackage{courier}        
	\usepackage[usenames,dvipsnames,svgnames]{xcolor}						
\usepackage{makeidx}         
\usepackage{graphicx}        
\usepackage{multicol}        
\usepackage[bottom]{footmisc}
\usepackage[square,numbers]{natbib} 

\usepackage[colorlinks=true, linkcolor=DarkBlue, citecolor=DarkBlue, urlcolor=DarkBlue, plainpages=false, pdfpagelabels, hypertexnames=false, unicode]{hyperref}

\hypersetup{linkcolor=DarkBlue}

\makeindex

\def\xT{\textbf{x}_T }

\newcommand{\feq}{f_{\rm eq}}

\newcommand{\teq}{\tau_{\rm eq}}

\newcommand{\eq}{{\rm eq}}

\newcommand{\vp}{\vphantom{\frac{}{}}}
\newcommand{\dS}{d^3\!\Sigma}   
\newcommand{\dP}{d^3\! p}  
\newcommand{\e}{{\cal E}}  
\newcommand{\p}{{\cal P}}  
\newcommand{\n}{{\cal N}}

\newcommand{\SHARE}{\texttt{SHARE\,}}
\newcommand{\THERMINATOR}{\texttt{THERMINATOR\,}}
\newcommand{\THERMINATORTWO}{\texttt{THERMINATOR2\,}}
 
\begin{document}
\title*{Monte-Carlo statistical hadronization in relativistic heavy-ion collisions}
\titlerunning{Monte-Carlo statistical hadronization in relativistic heavy-ion collisions} 
\author{Radoslaw Ryblewski}
\authorrunning{R.~Ryblewski}  
\institute{Radoslaw Ryblewski \at Institute of Nuclear Physics, Polish Academy of Sciences, PL-31342 Krak\'ow, Poland,\\ 
\email{Radoslaw.Ryblewski@ifj.edu.pl}
} 
\maketitle
\abstract{A brief introduction to the statistical hadronization approach to particle production in relativistic heavy-ion collisions is given. In the context of fluid dynamics modeling various aspects of hadron emission at the freeze-out are discussed. Practical applications of the presented concepts are presented within the \THERMINATOR~Monte-Carlo hadron generator.}
%
\section{Relativistic heavy-ion collisions}
\label{sec:1}
\sectionmark{Relativistic heavy-ion collision}
%
The relativistic heavy-ion collision experiments are a perfect tool for studying, in a controlled and reproducible manner, the properties of strongly interacting matter at high energies. Unlike the collisions of more elementary particles, they provide unique opportunity to reach the thermodynamic equilibrium needed for investigating the phase diagram and transport properties of the QCD matter. Assuming that (local) thermal equilibrium is achieved promptly, and that the interactions are strong enough to maintain this state throughout subsequent evolution, the expansion of such a system should, in principle, follow the laws of  relativistic fluid dynamics  \cite{Yan:2017ivm} \footnote{Note that several recent studies indicate that fluid dynamics may be applicable also in the situations where the produced system is locally far off equilibrium \cite{Florkowski:2010cf,Martinez:2010sc,Alqahtani:2017jwl,Strickland:2014pga,Alqahtani:2017mhy}, see also \cite{Florkowski:2013lya,Florkowski:2013lza,Denicol:2014xca,Denicol:2014tha}.}. Given initial conditions for initialization of fluid dynamical fields, and the prescription for the hadron emission from the fluid, the fluid dynamics provides straightforward and intuitive way to study properties of the produced matter, as encoded in its equation of state.  Although still some important questions remain, \emph{e.g.}, concerning the formulation of the theory itself \cite{Florkowski:2017olj} (in particular related to the form of the transport coefficients \cite{Jaiswal:2014isa,Florkowski:2015lra,Tinti:2016bav}), the successes of the models employing fluid dynamics concepts already have shown  (almost)-perfect fluidity of the quark-gluon plasma and established a sort of hydro-like ``Standard Model'' of heavy-ion collisions. While recently a significant progress in the  determination of the initial state of heavy-ion collisions has been made, using, \emph{i.a.}, non-equilibrium effective field theory and gauge/gravity correspondence, the hadronic production from such a system is still poorly understood. Huge majority of the approaches use (to some level heuristic, yet surprisingly successful) prescriptions for the hadronization process dating back to times of Fermi, Landau, and Hagedorn. 

In these lectures we will briefly review some of the concepts of particle decoupling and statistical hadronization as applied to heavy-ion collisions showing, in the end, their remarkable efficacy in describing some of experimentally observed phenomena.

In this work the we use natural units where $c=k_B=\hbar =1$. The bold font denotes the vectors in the transverse $x-y$ plane.
%
\section{Relativistic perfect fluid dynamics}
\label{sec:2a}
\sectionmark{Relativistic perfect fluid dynamics}
%
The simplest and, at the same time, the only unambiguously~\footnote{Some sort of ambiguity arises in the case when dissipative effects are present in the system. In such a situation additional assumptions on the evolution of dissipative quantities (\emph{e.g.} shear stress tensor and bulk viscous pressure) are required, resulting in additional equations of motion. The latter may differ significantly in various approaches~\cite{Florkowski:2016kjj}.} derivable relativistic fluid dynamical equations are that of \emph{relativistic perfect fluid dynamics}~\cite{Landau:1959,Misner:1974qy,deGroot:1980,rezzolla2013relativistic}. Due to their simplicity they are extensively applied to various systems in physics, including the evolution of strongly interacting matter produced in relativistic heavy-ion collisions. Although the literature on the subject is rather extensive (see \emph{i.e.} Refs.~\cite{Florkowski:2017olj,Stoecker:1986ci,Rischke1999,Kolb:2003dz,Huovinen:2006jp,Florkowski:2010zz,Romatschke:2009im,Gale:2013da,Jaiswal:2016hex} and the references therein), we will review herein its basic aspects to set the stage for further discussion. 

The equations of relativistic perfect fluid dynamics for the (net) charge-free matter follow solely from the local conservation laws of energy and momentum~\cite{Landau:1959,Misner:1974qy,deGroot:1980,rezzolla2013relativistic}, which in the Minkowski coordinates  may be formulated in the following covariant form \footnote{In the case of curvilinear coordinates, even if the space-time is considered to be flat (in the sense of globally vanishing Riemann tensor), one should replace the partial derivative $\partial_\mu=(\partial_t, - \nabla)$ in  Eq.~(\ref{emc}) with \emph{covariant derivative} $d_\mu$.}
\begin{equation}
\partial_\mu T^{\mu\nu}(x) = 0,
\label{emc}
\end{equation}
where $T^{\mu\nu}(x)$ is the energy-momentum tensor and $x^\mu = (t, \xT, z)$.

In addition, for a multicomponent system which possesses $N$ conserved charges $Q_i$ one should supplement Eq.~(\ref{emc}) with $N$ continuity equations for the respective charge currents $N_i^\mu$  
\begin{equation}
\partial_\mu N_i^{\mu}(x) = 0  \qquad \left(i=1\dots \,N\right).
\label{nc}
\end{equation}
One may consider, \emph{i.e.}, $Q_i=\{B, I_3, S, C\}$, where $B, I_3, S$ and $C$ denote baryon number, third component of the isospin, strangeness and charm, respectively \footnote{Equivalently, instead of the third component of the isospin one may use the electric charge.}.

The main assumption defining the perfect fluid is that each fluid element, when considered in its local rest frame (LRF), is exactly in \emph{thermal and chemical equilibrium} state. This is expressed by the static equilibrium (isotropic) form of the energy-momentum tensor 
\begin{equation}
 T^{\mu\nu}_{\rm LRF} (x) =  {\rm diag}\left(\vp \e(x), \p(x), \p(x), \p(x) \right),
\label{emtLRF}
\end{equation}
where $\e(x)$ and $\p(x)$ denote the equilibrium energy density and pressure, respectively. One may also easily convince oneself that in the perfect fluid case the charge currents must all have the following LRF form
\begin{equation}
N^\mu_{i, {\rm LRF}} (x) =  \left(\vp\!\!\n_i(x), 0, 0, 0 \right),
\label{ncLRF}
\end{equation}
where $\n_i(x)$ represents the density of the charge $Q_i$; otherwise dissipative effects have to occur.

In general (laboratory) frame each fluid element moves with a fluid four-velocity $u^\mu (x) \equiv \gamma \left(1, \textbf{v}_T, v_z\right)$,  satisfying normalization condition  $u^\mu u_\mu=1$. The form of the energy-momentum tensor in this frame can be obtained by applying a general (canonical) Lorentz boost transformation  to Eq.~(\ref{emtLRF})
\begin{equation}
T^{\mu\nu}(x) = \Lambda^\mu_{\,\,\,\alpha}\ (u^\lambda)\,\Lambda^\nu_{\,\,\,\beta} (u^\lambda)\,T_{\rm LRF}^{\alpha\beta}(x),
\label{transform}
\end{equation}
 where the boost matrix $\Lambda^\mu_{\,\,\,\nu}$ is defined as follows
\begin{equation}   
\Lambda^{\mu}_{ \,\,\,\nu}(u^{\lambda}) \equiv \left(
\begin{array}{rrrr}
\gamma &  -\gamma v_x  &  -\gamma v_y  &  -\gamma v_z \\
-\gamma v_x & 1 + (\gamma - 1) \frac{v_x^2}{v^2}    &     (\gamma - 1) \frac{v_x v_y}{v^2}  &     (\gamma - 1) \frac{v_x v_z}{v^2} \\
-\gamma v_y &     (\gamma - 1) \frac{v_x v_y}{v^2}  & 1 + (\gamma - 1) \frac{v_y^2}{v^2}    &     (\gamma - 1) \frac{v_y v_z}{v^2} \\
-\gamma v_z &     (\gamma - 1) \frac{v_x v_z}{v^2}  &     (\gamma - 1) \frac{v_y v_z}{v^2}  & 1 + (\gamma - 1) \frac{v_z^2}{v^2}
\end{array} \right).\nonumber
\label{genboost}
\end{equation} 
Using covariant notation the result of Eq.~(\ref{transform}) may be expressed in the following form
\begin{eqnarray}
T^{\mu \nu}   =\e  u^\mu u^\nu -  \p\Delta^{\mu\nu},
\label{emt}
\end{eqnarray} 
where we introduced the symmetric projection operator on the space orthogonal to the fluid four-velocity, $\Delta^{\mu\nu} \equiv g^{\mu\nu} - u^\mu u^\nu$, which satisfies conditions $u_{\mu} \Delta^{\mu\nu} = 0$, $\Delta^{\mu}_{\,\,\,\alpha}\Delta^{\alpha\nu}=\Delta^{\mu\nu}$ and $\Delta^{\mu}_{\,\,\,\mu}=3$, and $g^{\mu\nu} = {\rm diag} \left(1,-1,-1,-1\right)$ is the metric tensor. 

Similarly, the Lorentz boost transformation applied to Eq.~(\ref{ncLRF}) leads to the following tensor decomposition of $N_i^\mu$ in the general frame
\begin{equation}
 N_i^{\mu}  =  \n_i \,u^\mu,
\label{pflux}
\end{equation}
so that in the LRF, where $u^\mu_{\rm LRF} =\left(1, \textbf{0}, 0\right)$, one has $\n_i=N_i^\mu u_{\mu, \rm LRF}$. 

Equation (\ref{emc}) may be rewritten in a somewhat more familiar form using Eq.~(\ref{emt}) and performing   projections perpendicular and parallel to the fluid four-velocity  
\begin{eqnarray} 
\label{emc1}
\Delta^\alpha_{\,\,\,\nu}  \partial_\mu T^{\mu \nu}  = (\e+\p)D u^\alpha-\nabla^\alpha \p &=&0, \\
\label{emc2}
u_\nu \partial_\mu T^{\mu \nu}= D\e + (\e+\p)\theta &=&0,
\end{eqnarray}
where $D\equiv u^\mu \partial_\mu$ is the co-moving time derivative, $\nabla^\mu \equiv \Delta^{\mu \alpha} \partial_\alpha$ is the spacial gradient, and $\theta\equiv\partial_\mu u^\mu$ is the expansion scalar. Equations (\ref{emc1})-(\ref{emc2}) are relativistic analogs of the Euler and continuity equations, respectively. Similarly, putting decompositions (\ref{pflux}) in Eqs.~(\ref{nc}) yields
\begin{eqnarray} 
\label{nc2}
 \partial_\mu N^{\mu}_i  =  D \n_i+ \n_i \theta&=&0 .
\end{eqnarray}

\begin{figure}[t] 
\begin{center}
\includegraphics[angle=0,width=0.6 \textwidth]{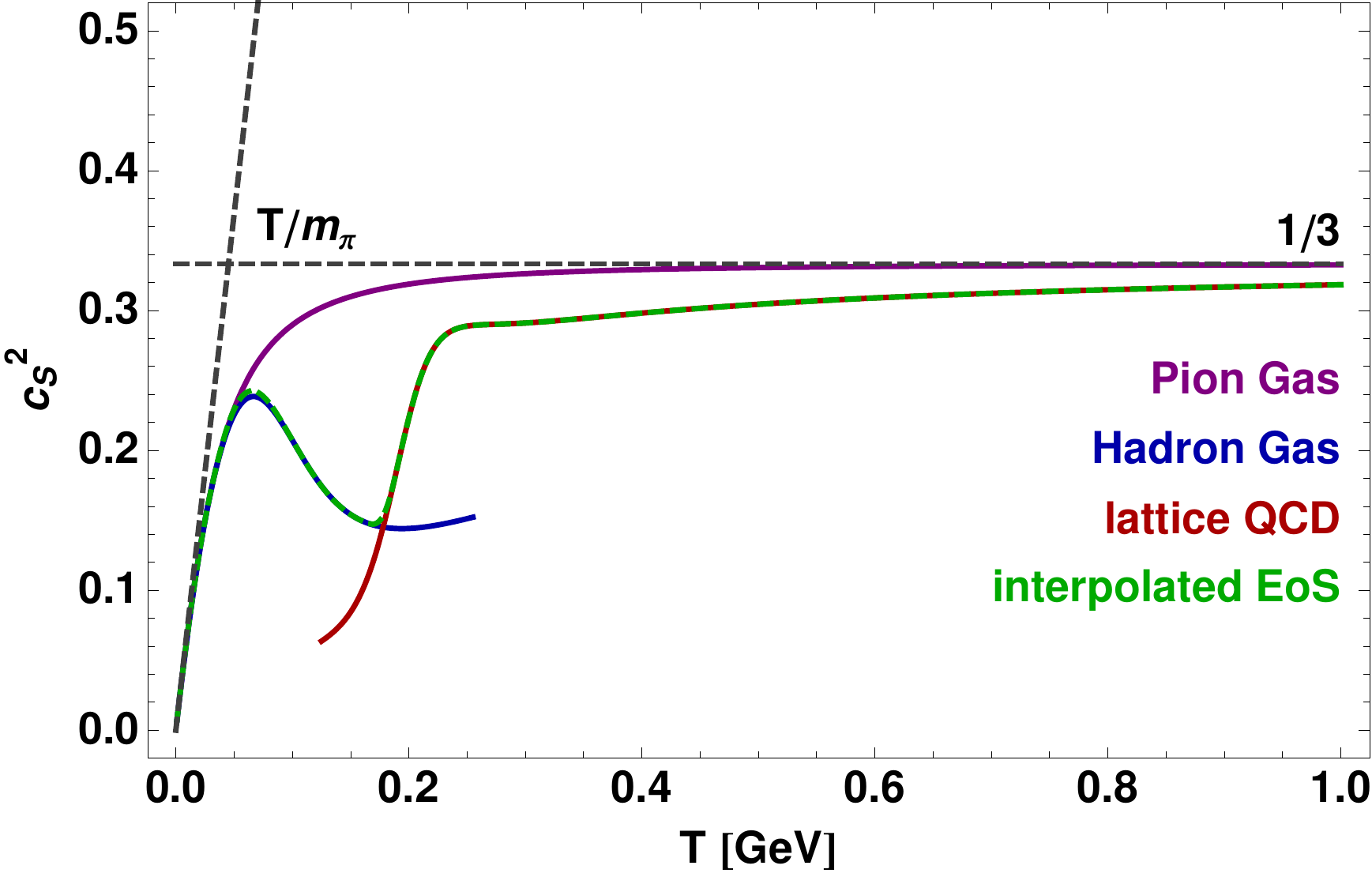}
\end{center} 
 \caption{Temperature dependence of the squared speed of sound $c_s^2=d\p/d\e$ for the hadron gas, lattice QCD quark-gluon plasma, and interpolation thereof as found in Ref.~\cite{Chojnacki:2007jc}.}
 \label{fig:eos}
\end{figure}

Equations (\ref{emc1})-(\ref{nc2}) contain together $4+N$ independent partial differential equations for the space-time evolution of $5+N$ quantities (three components of four-velocity, energy density, pressure and $N$ charge densities). In order for the system to be closed one has to introduce a material-specific \emph{equation of state} relating the pressure, the energy density and the charge densities in the system, $\p=\p(\e,\n_i)$. 
Since the system is locally in equilibrium such a relation exists and has to follow from the underlying microscopic theory describing the system. Henceforth, we will assume that the system created in heavy-ion collisions is charge-free, $\n_i(x) \equiv0$, which is a reasonable assumption for central rapidity region at the ultra-relativistic energies. The energy density and pressure for the charge-free matter in equilibrium may be directly related to the temperature of the system, $\e=\e(T)$, $\p=\p(T)$, see Sec.~\ref{sec:2c}. A number of studies show that the successful description of the experimental data requires the use of ``cross-over''-type equation of state of strongly interacting matter with the transition from the quark-gluon plasma phase to the hadron gas phase.  For the numerical results presented in the remaining part of these lectures we will use the results of the Ref.~\cite{Chojnacki:2007jc}. The temperature dependence of the square speed of sound $c_s^2 (T)=d\p/d\e$ obtained in Ref.~\cite{Chojnacki:2007jc} is shown in Fig.~\ref{fig:eos}  \footnote{Note that in the case of dissipative fluid dynamics due to the existence of transport coefficients the inclusion of the equation of state  is usually more involved \cite{Tinti:2016bav}.}. 

\begin{figure} 
\begin{center}
\includegraphics[angle=0,width=0.67 \textwidth]{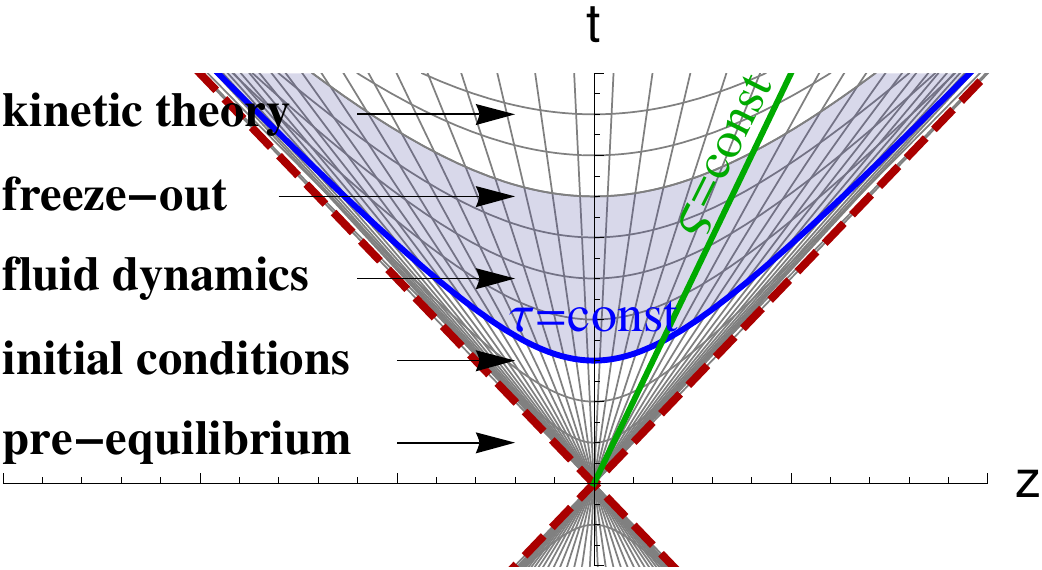}
\end{center} 
 \caption{The Milne coordinates shown in the Minkowski space-time. Directions $x$ and $y$ are suppressed.}
 \label{fig:milne}
\end{figure}
 
Except for quite limited number of special cases of highly-symmetric flow patterns, such as the Bjorken \cite{Bjorken:1982qr} or Gubser \cite{Gubser:2010ze} flows, Eqs. (\ref{emc1})-(\ref{nc2}) have to be solved numerically. In the case of modeling the collisions at relativistic energies, in which case the system is approximately boost-invariant (with respect to Lorentz boosts along the beam ($z$) direction) in the central rapidity region, the hydrodynamic evolution is preferably performed in \emph{Milne coordinates} $x^\mu = (\tau,\xT,\varsigma)$ instead of Minkowski coordinates $\tilde{x}^\mu = (t,\xT,z)$ 
\cite{Bjorken:1982qr}. The relation between the two is given by the following coordinate transformation
\begin{eqnarray}  
t&=&\tau \cosh \varsigma,\\
z&=& \tau \sinh \varsigma,
\label{coord}
\end{eqnarray}
with  $\tau= \sqrt{t^2-z^2}$ and $\varsigma = \tanh^{-1} \left(z/t\right)$ denoting longitudinal proper time and space-time rapidity, respectively; see Fig.~\ref{fig:milne}. 
 
In this coordinate system it is also convenient to parametrize the fluid four-velocity in the following form
\begin{equation} 
u^\mu = \left(u_0 \cosh {\rm y}_u, \textbf{u}_T, u_0 \sinh {\rm y}_u \right),   
\label{umilne}  
\end{equation}
where ${\rm y}_u$ is the longitudinal rapidity of the fluid, and $u_0 = \sqrt{1+u_T^2}$, with $u_T=\sqrt{u_x^2+u_y^2}$.
%
\section{Fluid dynamics from kinetic theory}
\label{sec:2c}
\sectionmark{Fluid dynamics from kinetic theory}
%
It is instructive to see the relation between the fluid dynamics, as a general classical field theory, and the relativistic kinetic theory. The latter is based on the knowledge of the \emph{single-particle distribution function} $f(x,p)$, which is defined through the number of (on-shell) particles $d N$ in the phase-space volume $d^3 \!x\,\dP$ located at the phase-space point $(x^\mu, p^\mu)$, where $p^\mu=(E_p, \textbf{p}, p_z)$ with $E_p=\sqrt{m^2+\textbf{p}^2+p_z^2}$. The evolution of $f(x,p)$ follows from the standard relativistic Boltzmann equation
\begin{equation}
p^\alpha \partial_\alpha f=-C[f],
\label{BE}
\end{equation}
where $C[f]$ is the collisional kernel, which, in general, may have highly complicated form. In global equilibrium $f(x,p)$ is stationary, which means that $C[f]$ vanishes in two very different regimes: free-streaming (no interactions) and equilibrium (strongest possible interactions). Often the collisional kernel is treated in the relaxation-time approximation, 
\begin{equation}
C[f]=p_\mu u^\mu\frac{ f-\feq  }{\teq},
\label{ccRTA}
\end{equation}
where $\teq$ is the relaxation time, and $\feq$ is the  equilibrium distribution.

Equations of motion for the soft modes of the system, identified with the fluid dynamical sector of the theory, may be derived by taking the lowest-$n$ momentum moments \cite{Denicol:2014loa,Strickland:2014pga,Alqahtani:2017mhy}, 
\begin{equation}
\hat{{\cal I}}^{\mu_1\cdots\mu_n}\equiv  \int \!dP\, p^{\mu_1}p^{\mu_2}\cdots p^{\mu_n}, \qquad \qquad 
\int dP \equiv  \int \frac{\dP}{E_p},
\end{equation}
of the Boltzmann equation (\ref{BE}), which gives
\begin{equation}  
\partial_{\alpha} {\cal I} ^{\alpha \mu_1\cdots\mu_n}= -{\cal C}^{\mu_1\cdots\mu_n}[f],
\label{EOM} 
\end{equation}
where we defined 
\begin{eqnarray} 
{\cal I} ^{\alpha\mu_1\cdots\mu_n} &\equiv& \hat{{\cal I}} ^{\alpha\mu_1\cdots\mu_n} f ,
\label{mom1} \\{\cal C}^{\mu_1\cdots\mu_n}[f] &\equiv& \hat{{\cal I}} ^{\mu_1\cdots\mu_n} C[f].
\label{mom2} 
\end{eqnarray} 
Explicitly, the first two moments  lead  to the following set of dynamical equations, 
\begin{equation}
\partial_\mu {\cal I} ^\mu = u_\mu \frac{{\cal I}^\mu_\eq-{\cal I} ^\mu}{\teq},
\label{ncK}
\end{equation}
\begin{equation} 
\partial_\mu {\cal I} ^{\mu\nu}=u_\mu \frac{{\cal I}^{\mu\nu}_\eq-{\cal I} ^{\mu\nu}}{\teq}.
\label{emcK}
\end{equation}
One may immediately identify zeroth and first moments of the distribution function as particle four-current and the energy-momentum tensor,
\begin{eqnarray} 
N^\mu &\equiv& {\cal I}^{\mu}
\label{ident1} \\T^{\mu\nu} &\equiv& {\cal I}^{\mu\nu}.
\label{ident2} 
\end{eqnarray} 
The conservation of particle current $u_\mu ({\cal I}^\mu_\eq-{\cal I} ^\mu)=0$, and energy-momentum tensor $u_\mu ({\cal I}^{\mu\nu}_\eq-{\cal I} ^{\mu\nu})=0$ thus leads to Eqs.~(\ref{emc})-(\ref{nc}).

Using decompositions of the particle four-current (\ref{pflux}) and the energy-momentum tensor (\ref{emt}), and the knowledge of the LTE distribution function $f_{\rm eq} = f(p_\mu u^\mu, T, \mu_{\rm i})$, see Eq.~(\ref{eqdistr}) in Sec.~\ref{sec:6}, one may find explicit forms of the thermodynamic variables, $\e=\e(T,\mu_{\rm i})$, $\p=\p(T,\mu_{\rm i})$, and $\n=\n(T,\mu_{\rm i})$. The latter define the equation of state of the system within kinetic theory. For the conformal charge-free system one gets $\e(T) =3\p(T) = 3 T \n (T) \sim T^4$ \cite{Florkowski:2010zz}.
%
\section{Event-averaged initial conditions for fluid dynamics}
\label{sec:2b}
\sectionmark{Initial conditions for fluid dynamics}
%
In general, Eqs.~(\ref{emc1})-(\ref{nc2}) have to be supplemented with proper \emph{initial conditions}, specified on the hypersurface of constant longitudinal proper time $\tau=\tau_{\rm i}$, which usually  defines beginning of the fluid dynamical evolution. In particular, one has to provide  $\e(\tau_{\rm i},\xT,\varsigma)$, $u_x(\tau_{\rm i}, \xT,\varsigma)$, $u_y(\tau_{\rm i}, \xT,\varsigma)$, $y_u(\tau_{\rm i},\xT,\varsigma)$, and $\n_i(\tau_{\rm i},\xT,\varsigma)$. These should follow from some microscopic models of the initial state created in heavy-ion collision, though usually they are, to some extent, just fitted to reproduce the data. 
%
\begin{figure} [t]
\begin{center}
\includegraphics[angle=0,width=0.67 \textwidth]{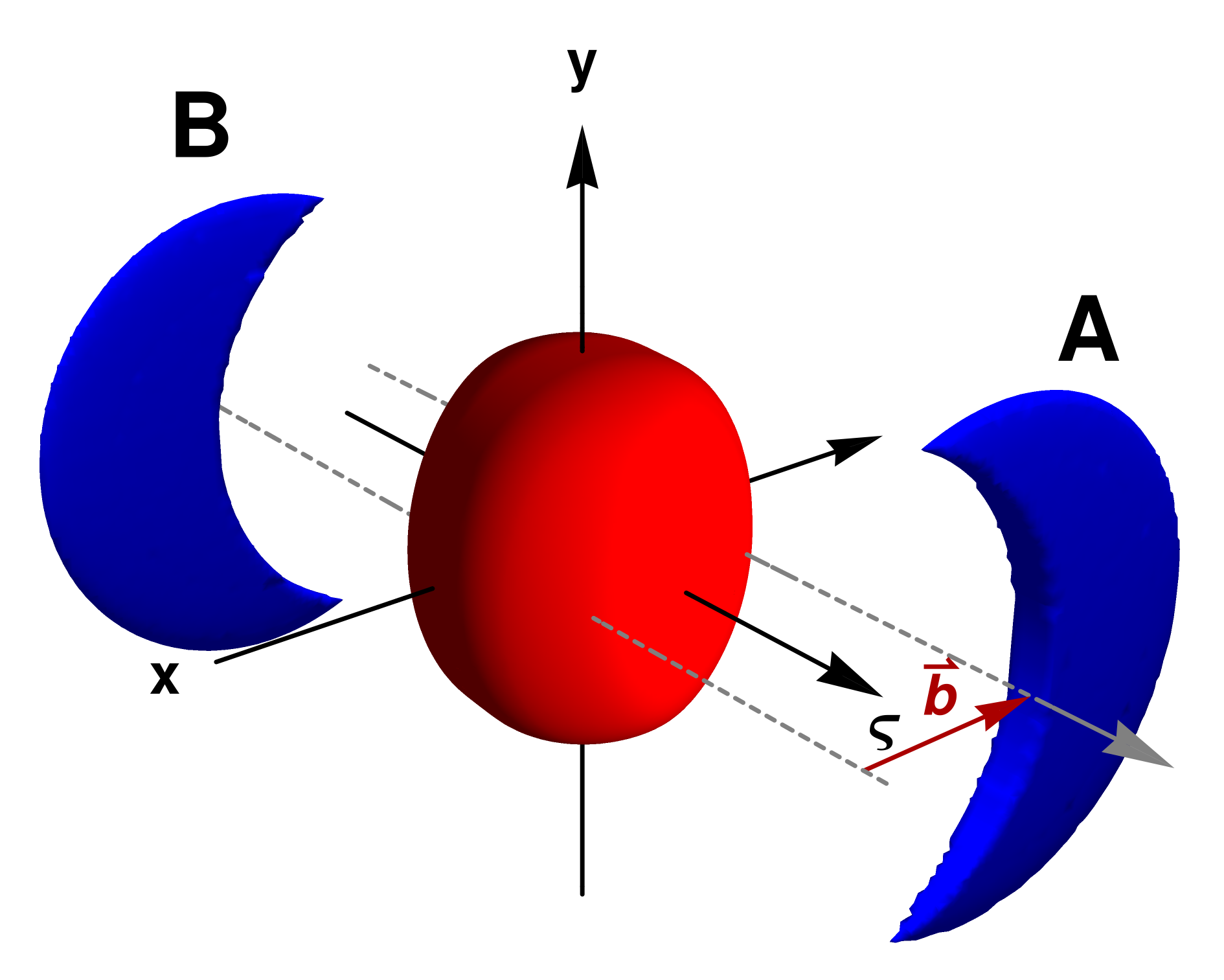}
\end{center} 
 \caption{The geometry of heavy-ion collision in the Milne coordinates.}
 \label{fig:coll}
\end{figure}
 
For the initial energy density profile, we will use the \emph{tilted source} model \cite{Bozek:2010bi} of the initial state, which was applied quite successfully to describe various experimental observables measured at RHIC, including the so called directed flow $v_1$ component of the Fourier decomposition of the azimuthal particle spectra. The initial energy density profile within this model is proportional to the density of sources $n({\bf x}_T,\varsigma,b)$ 
\begin{equation}
 \e(\tau_{\rm i},{\bf x}_T, \varsigma,b) = 
  \e_{\rm i} \, \frac{ n({\bf x}_T,\varsigma,b)}{n(\textbf{0},0,0)} ,
\label{sig2}
\end{equation}
where 
\begin{equation}
n({\bf x}_T,\varsigma,b) = G(\varsigma)\left\{(1-\kappa)  \Big[ W_A({\bf x}_T,b) F(\varsigma) + W_B({\bf x}_T,b) F(-\varsigma)\Big] + \kappa   B({\bf x}_T,b) \right\}.
\label{denofsourc}
\end{equation}  
 
The   functions $W_{A(B)}$, and $B$ are the density of wounded nucleons from the  nucleus A (B), and the density of binary collisions, respectively, both specified at a certain value of the impact parameter $b=|\textbf{b}|$, see Fig.~\ref{fig:coll}. These quantities are determined entirely from the optical limit of the \emph{Glauber model} \cite{Miller:2007ri}. The admixture of binary collisions is controlled by the parameter $\kappa =0.14$ that is typically fitted to reproduce the centrality dependence of charge hadron multiplicity.  As herein we assume that the system is charge-free, $\n_i(\tau_{\rm i}, x,y,\varsigma)\equiv 0$, the initial central energy density $\e_{\rm i}$ of the system may be translated to its initial central temperature $T_{\rm i}=T_{\rm i}(\e_{\rm i})$, which is fitted to reproduce the total number of charged particles produced in the experiment.

%
\begin{figure}[t]
\begin{center} 
\includegraphics[angle=0,width=0.7 \textwidth]{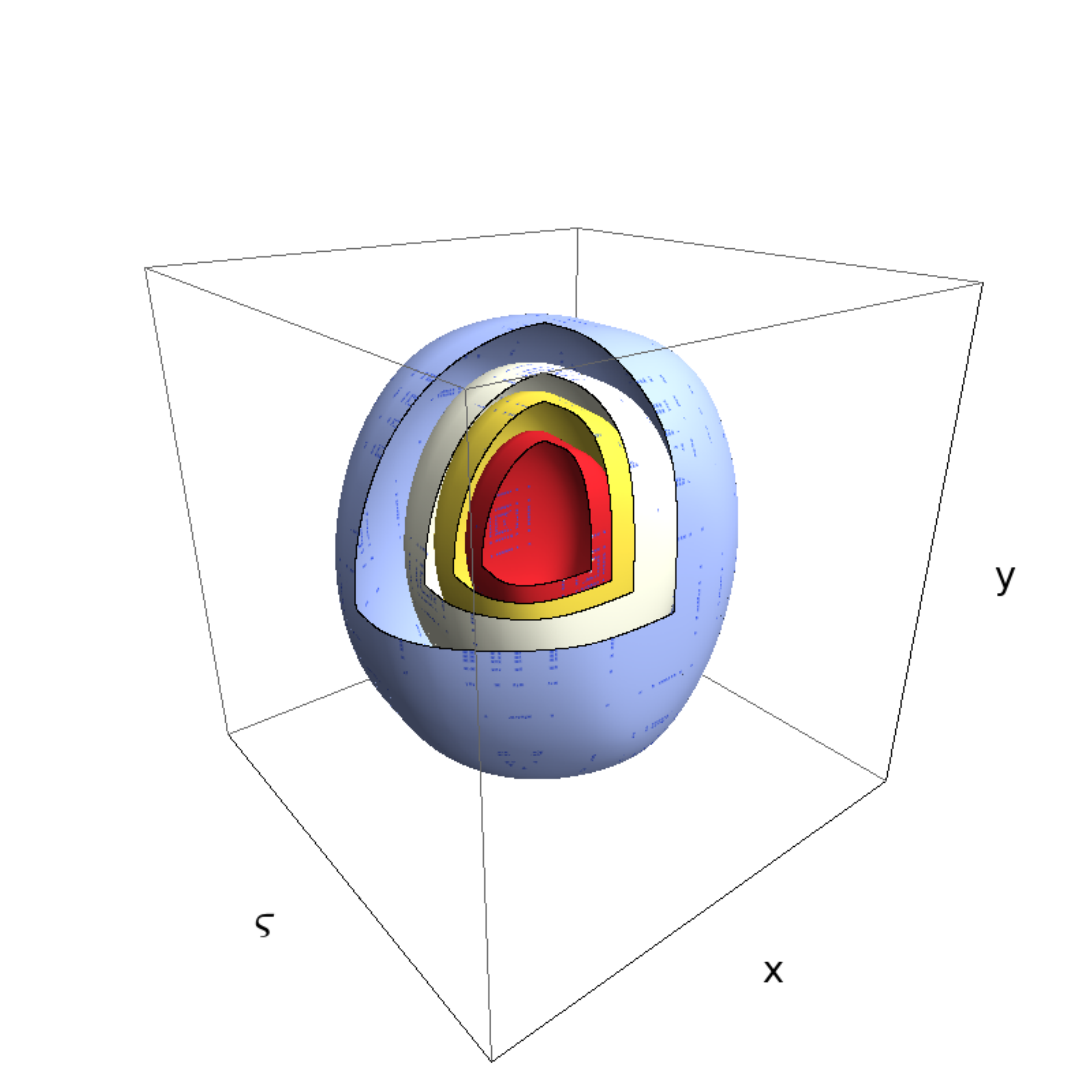}\\
\includegraphics[angle=0,width=0.47 \textwidth]{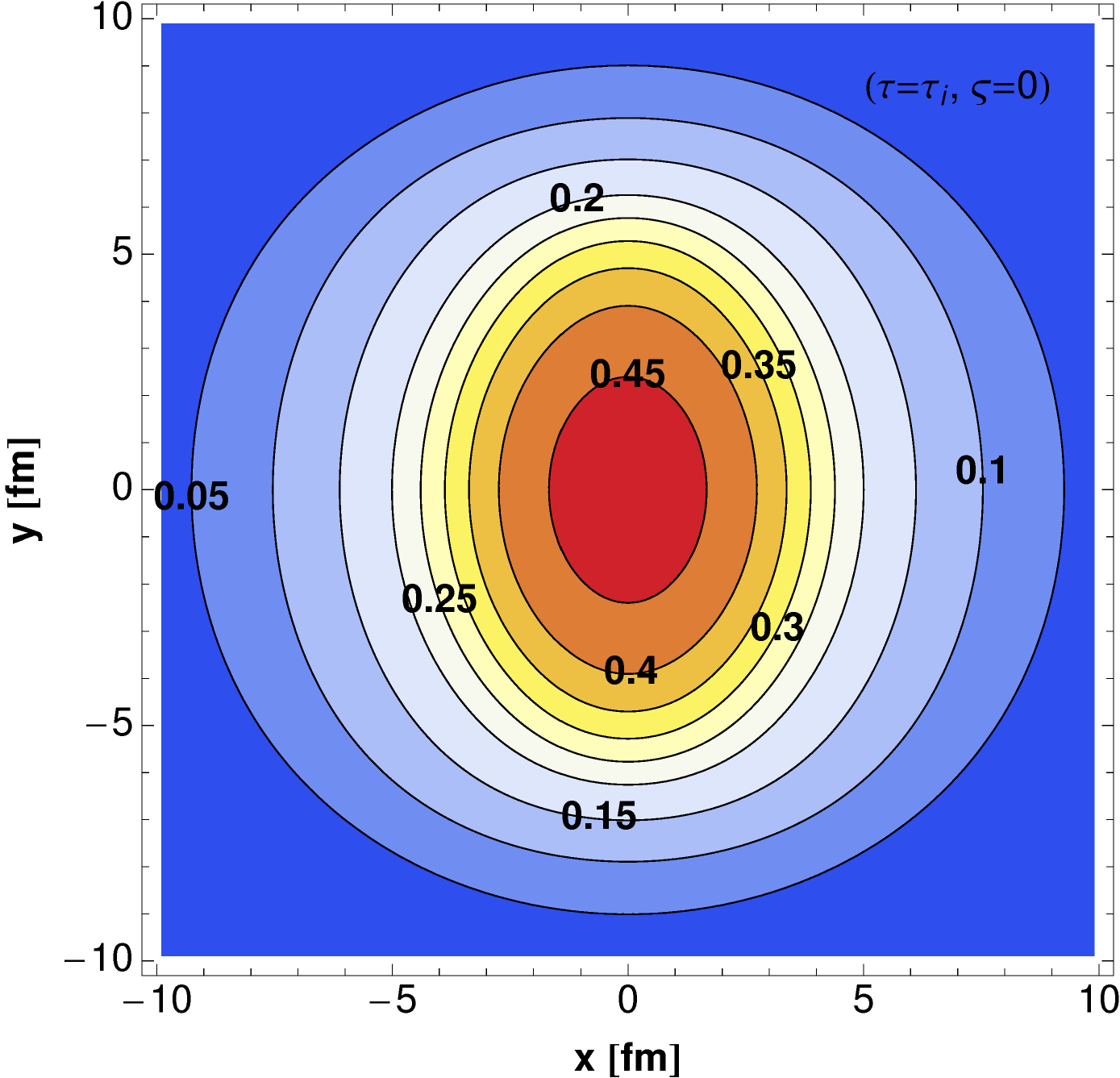}
\includegraphics[angle=0,width=0.47 \textwidth]{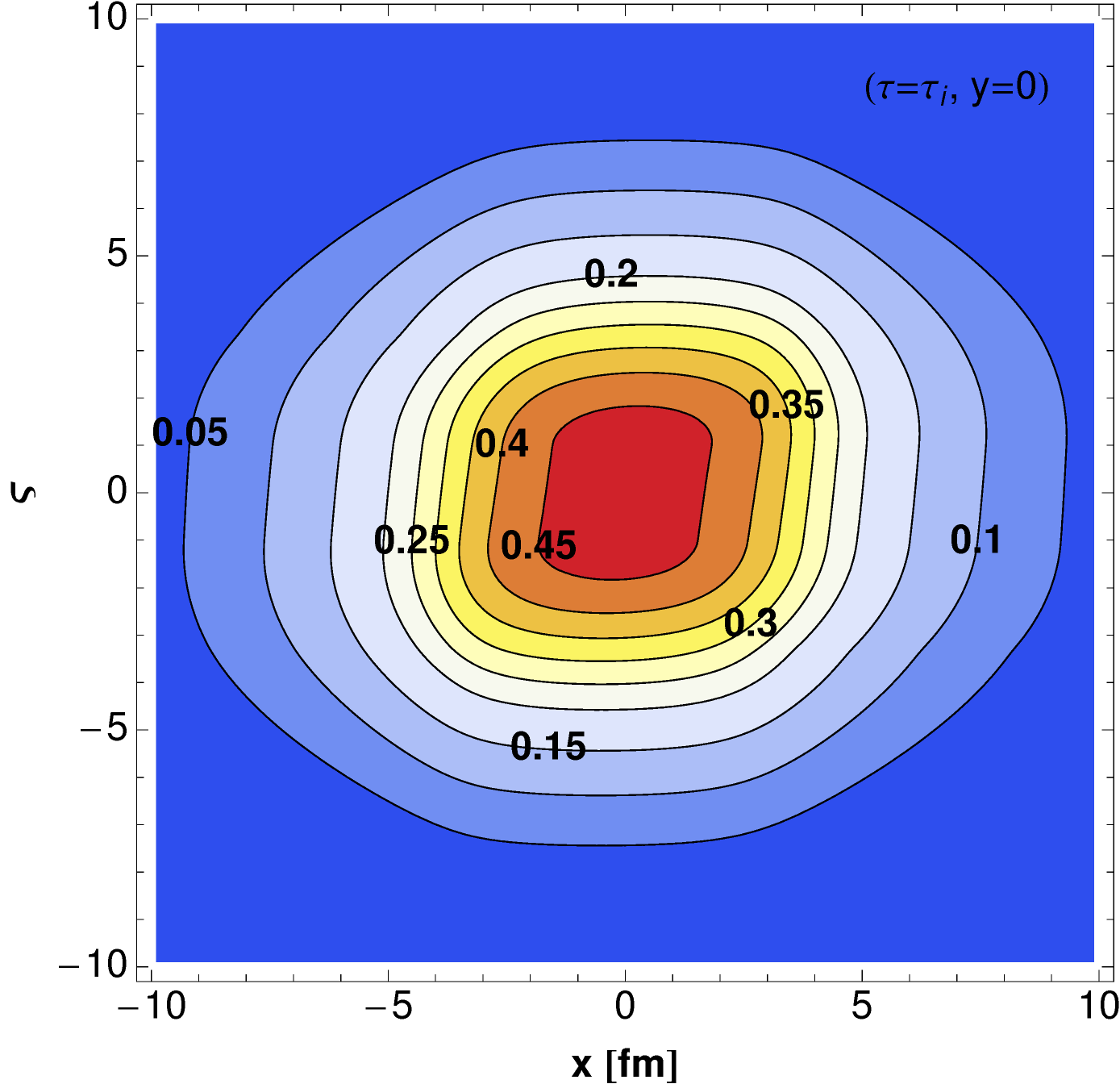}
\end{center} 
 \caption{(Top panel) The isothermal surfaces $(T\in\{0.4,0.3,0.2,0.1\} {\rm \,GeV})$ of the tilted source in the Milne coordinates. (Bottom panels) The isothermal contours of the initial temperature profile of fluid dynamic evolution in the Milne coordinates in $x-y$ (left) and $x-\varsigma$ (right) plane.}
 \label{fig:ini}
\end{figure}
%
The functional form of the density profile in rapidity in Eq.~(\ref{denofsourc}) is 
\begin{equation}
G(\varsigma) \equiv \exp \left[ - \frac{(\varsigma - \Delta \varsigma)^2}{ 2 \sigma_\varsigma^2 } \, \Theta (|\varsigma| - \Delta \varsigma) \right] \, .
\label{eq:rhofunc}
\end{equation}
The parameters in Eq.~(\ref{eq:rhofunc}) are deduced from the fits to the final rapidity spectrum of charged hadrons. For RHIC the fit results in  $\Delta\varsigma = 2.3$ and $\sigma_{\varsigma} = 1.6$. The tilt of the source results from the preferred particle emission from the moving participant nucleon into its forward hemisphere, and may be parametrized as follows \cite{Bozek:2010bi}
\begin{eqnarray}
F(\varsigma) =
\left\{ \begin{array}{lcccc}
0  & \,\,\,\,\,\,\,\,\,\,\,\,\,\,\,\ & \mbox{ } & \,\,\,
& \varsigma < -y_{\rm N} \, ,
 \\ (\varsigma+y_{\rm N})/(2 y_{\rm N}) & & \mbox{if} &
& -y_{\rm N} \leq \varsigma \leq y_{\rm N}\, , \\
1 & & \mbox{ } & 
& \varsigma > y_{\rm N}\, ,
\end{array}\right. \,\,\,\,\,\,\,\,\,\,\,\,\,
\end{eqnarray} 
where $y_{\rm N} = \log(2\sqrt{s_{\rm NN}}/(m_{\rm N}))-\varsigma_{\rm shift}$ is the nucleon rapidity  shifted by the value $\varsigma_{\rm shift}=2$ (treated as a phenomenological parameter), $\sqrt{s_{\rm NN}}$ is the center-of-mass energy per nucleon pair, and $m_{\rm N}$ is the nucleon mass.
The resulting initial temperature profile in $x-y$ plane and $x-\varsigma$ plane is shown in bottom panels of Fig.~\ref{fig:ini}, respectively.

Finally, the flow in the transverse plane, as usual, is assumed to vanish initially, $u_x(\tau_{\rm i}, \xT,\varsigma) =0$, $u_y(\tau_{\rm i}, \xT,\varsigma)=0$,  and the flow in the longitudinal direction is assumed to have the Bjorken-type scaling form $y_u(\tau_{\rm i}, \xT,\varsigma)=\varsigma$.

\section{Particle decoupling}
\label{sec:3}
\sectionmark{Particle decoupling}
%
%
\begin{figure}[t]
\begin{center}
\includegraphics[angle=0,width=0.47 \textwidth]{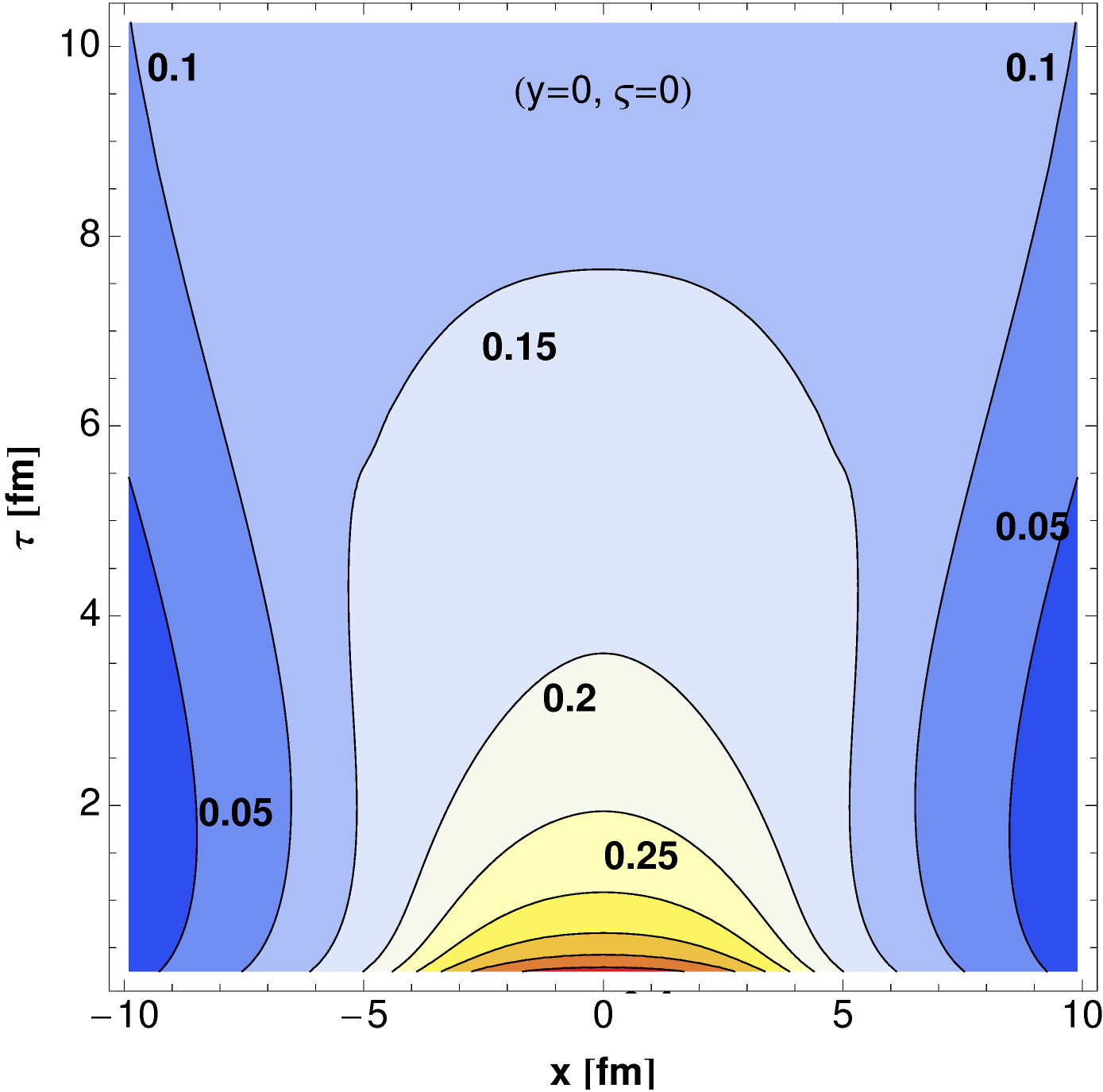}
\includegraphics[angle=0,width=0.47 \textwidth]{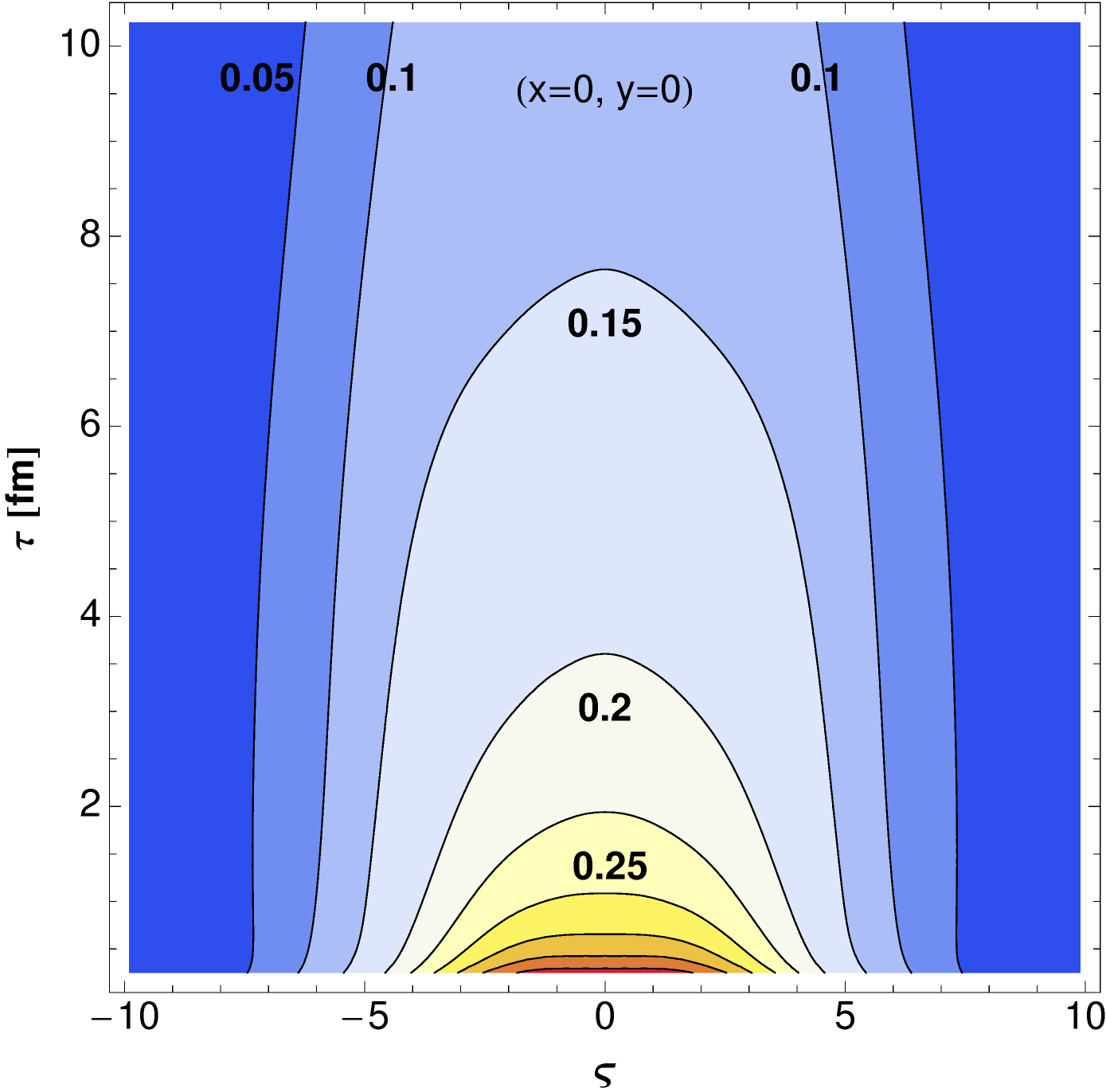}
\end{center} 
 \caption{The isothermal contours of the fluid dynamical evolution in the Milne coordinates.}
 \label{fig:evo}
\end{figure}
%
Relativistic perfect fluid dynamics describes, by definition, the infinitely strongly-coupled system of particles evolving from one local thermal equilibrium state to another. When applied to  relativistic heavy-ion collisions one immediately realizes that the latter assumption breaks down as the evolution proceeds. The rapid expansion of the created fireball into the vacuum leads to its cooling and dilution, see Fig.~\ref{fig:evo}. Eventually, the particle scatterings become too rare to prevent the particles from leaving the fluid.  As a result the local thermal equilibrium cannot be maintained anymore and the fluid description breaks down. This complicated gradual process of particle decoupling from the fluid is often called the \emph{freeze-out} \cite{Stoecker:1986ci,Rischke1999,Kolb:2003dz,Huovinen:2006jp,Florkowski:2010zz,Romatschke:2009im,Gale:2013da,Jaiswal:2016hex}. 

As the interactions cease and the system becomes rarefied the kinetic theory description in terms of hadronic degrees of freedom and scattering cross sections becomes more adequate. A possible way to describe this process is to compare locally the time scale of the expansion of the fluid $\tau_{\rm  exp}$ (which drives the system out of equilibrium), and the time scale characterizing collisions between the particles $\tau_{{\rm coll}}^k$ (which tend to restore it) \cite{Bondorf:1978kz}. In the differential form, the decoupling may be formulated as the following inequality \cite{Heinz:2007in}
\begin{eqnarray} 
\tau_{{\rm coll}}^{k} \ge   \tau_{\rm  exp} ,
\label{freezeout}
\end{eqnarray}
where $\tau_{\rm exp}\sim 1/\theta(x)$ (see Sec.~\ref{sec:2a}), and $\tau_{\rm coll}^{k}\sim 1/\sum\limits_l\langle \sigma_{kl} v_{kl} \rangle\, \tilde{\n}_l(x)$, with $\sigma_{kl}$ denoting the scattering cross section between the particle species $k$ and $l$, $v_{kl}$ being their relative velocity in the center of mass frame, and $\tilde{\n}_l(x)$ describing  respective particle densities. If the condition (\ref{freezeout}) is satisfied the particle species $k$ start to decouple from the fluid. 
 
A few comments are in order here:
\begin{itemize}
\item Both, the flow velocity $u^\mu(x)$ and particle densities $\tilde{\n}_l(x)$ are, in general, space-time dependent quantities, which means that the freeze-out process begins at different space-time points of the fluid.

\item In general, the cross sections $\sigma_{kl}$ depend on the particle species, thus some particle species   decouple ``before'' others. In the case of ultra-relativistic heavy-ion collisions the scattering cross section is usually dominated by a single species (pions), whose freeze-out triggers others. 

\item In perfect fluid dynamics, to which we restrict ourselves, the particle density $\tilde{\n}_l(x)\sim T^3(x)$. As a result, the condition (\ref{freezeout}) is usually significantly simplified to the condition of temperature dropping below a certain freeze-out temperature \cite{Rischke1999}
\begin{eqnarray} 
T_{\rm freeze} \ge   T(x).
\label{freezeoutT}
\end{eqnarray}

\item The total cross section $\sigma_{kl}=\sigma_{kl}^{\rm el}+\sigma_{kl}^{\rm in}$ is always larger than the elastic one $\sigma_{kl}^{\rm el}$, which implies that the particle-number changing processes cease before the momentum changing processes. This results in the distinction between the \emph{chemical freeze-out} (inelastic collisions stop), and the \emph{kinetic/thermal freeze-out} (elastic collisions stop). It means that, typically, chemical freeze-out takes place at higher temperature than the thermal one,
\begin{eqnarray} 
 T_{\rm  chem}  \ge   T_{\rm  therm}.
\label{freezeout2}
\end{eqnarray}

\item Equation (\ref{freezeout}) may, in general, involve   additional unknown parameter of the order of $1$, which sets the overall scale for the freeze-out processes \cite{Heinz:2007in}. 
\end{itemize} 
%
\section{\emph{Single-freeze-out} scenario}
\label{sec:4} 
%
The dynamical description of the particle decoupling according to Eq.~(\ref{freezeout}) is quite difficult to realize in practice.
Instead, usually a significant simplification of the freeze-out dynamics, often called the \emph{single-freeze-out} model, is adopted \cite{Broniowski:2001we,Broniowski:2001uk,Broniowski:2002nf,Broniowski:2002wp,Bozek:2003qi,Broniowski:2003ax,Kisiel:2006is}. The latter relies on the assumption that the chemical and thermal freeze-out occur simultaneously. Within this framework one assumes that once the temperature $T(x)$ in the fluid decreases locally below a certain value $T_{\rm freeze}$ all particle species  decouple  completely from the fluid \footnote{Although the assumption of isothermal freeze-out seems to be crude, it was shown to give a quite reasonable approximation of the differential freeze-out condition, Eq.~(\ref{freezeout}) \cite{Rischke1999,Kolb:2003dz}.}. Mathematically, the condition $T(x)=T_{\rm freeze}$ defines a three-dimensional freeze-out hypersurface $\Sigma$ in the four-dimensional Minkowski space-time ${\cal M}$ (see Sec.~\ref{sec:5}). The thickness of $\Sigma$ is idealistically assumed to be infinitesimal, which means that the freeze-out process takes place instantaneously. Just before crossing $\Sigma$, in the fluid phase, the matter is considered to be in local thermal and chemical equilibrium, so that phase-space distributions of the microscopic constituents follow the statistical ones (see Sec.~\ref{sec:6}). It is assumed that  freeze-out process itself, due to its instantaneous character, does not affect the phase-space distributions, so that equilibrium distributions are also shared by the particles emitted from the surface  $\Sigma$.

Outside the fluid region various approximations, usually based on some sort of transport theory, may apply. In the original single-freeze-out  model \cite{Broniowski:2001we,Broniowski:2001uk,Broniowski:2002nf} the particles created on $\Sigma$, termed as \emph{primordial}, are assumed to form \emph{ideal non-interacting gas of hadrons}, which undergo  \emph{free-streaming} to the detectors. However, the (primordial) hadron gas contains the full mass spectrum of hadronic resonances, which subsequently decay into \emph{stable} particles \footnote{The inverse processes, due to low probabilities, are usually neglected in this framework.}. Due to decays, the  abundances of stable particles, as well as their spectra, are modified (see Sec.~\ref{sec:9}). As a result, the temperature $T_{\rm freeze}$ should, in general, be interpreted as a \emph{switching temperature} (from fluid to particle description), rather than the true chemical and kinetic freeze-out temperature. Another possibility is to perform the switching at somewhat higher, often called \emph{particlization} \cite{Huovinen:2012is}, temperature $T_{\rm part}>T_{\rm freeze}$, when the system is considered to be still strongly interacting, but the transport theory description is more adequate.  The particle ensemble, together with their phase-space coordinates, is then passed to the quantum molecular transport models describing various phenomena in the dense hadronic medium. For the sake of simplicity, in this work we focus on the \emph{single-freeze-out} scenario solely. 

Finally, one should note here that it is commonly  assumed that the fluid dynamical description applies to the whole forward light-cone of the system and \emph{a posteriori} part of the space-time evolution of fluid satisfying $T(x)< T_{\rm freeze}$ is neglected and replaced with the transport theory description. This procedure obviously introduces some inconsistency at the boundary $\Sigma$, as outside the fluid regime the system is a non-interacting hadron gas (as opposed to the strongly-interacting fluid). Consequently, the boundary conditions at $\Sigma$ for the fluid evolution are different in the two cases. Possible consequences of these problems will be neglected herein. We will only note here, that these problems may be largely reduced when large anisotropies are included already within fluid dynamics stage which would describe its gradual break up \cite{Florkowski:2010cf,Martinez:2010sc,Alqahtani:2017jwl}. In this way the switching should introduce less uncertainty.
 
%
\section{Freeze-out hypersurface extraction}
\label{sec:5}
\sectionmark{Freeze-out hypersurface extraction}
%
%
\begin{figure}[t]
\begin{center}
\includegraphics[angle=0,width=0.6 \textwidth]{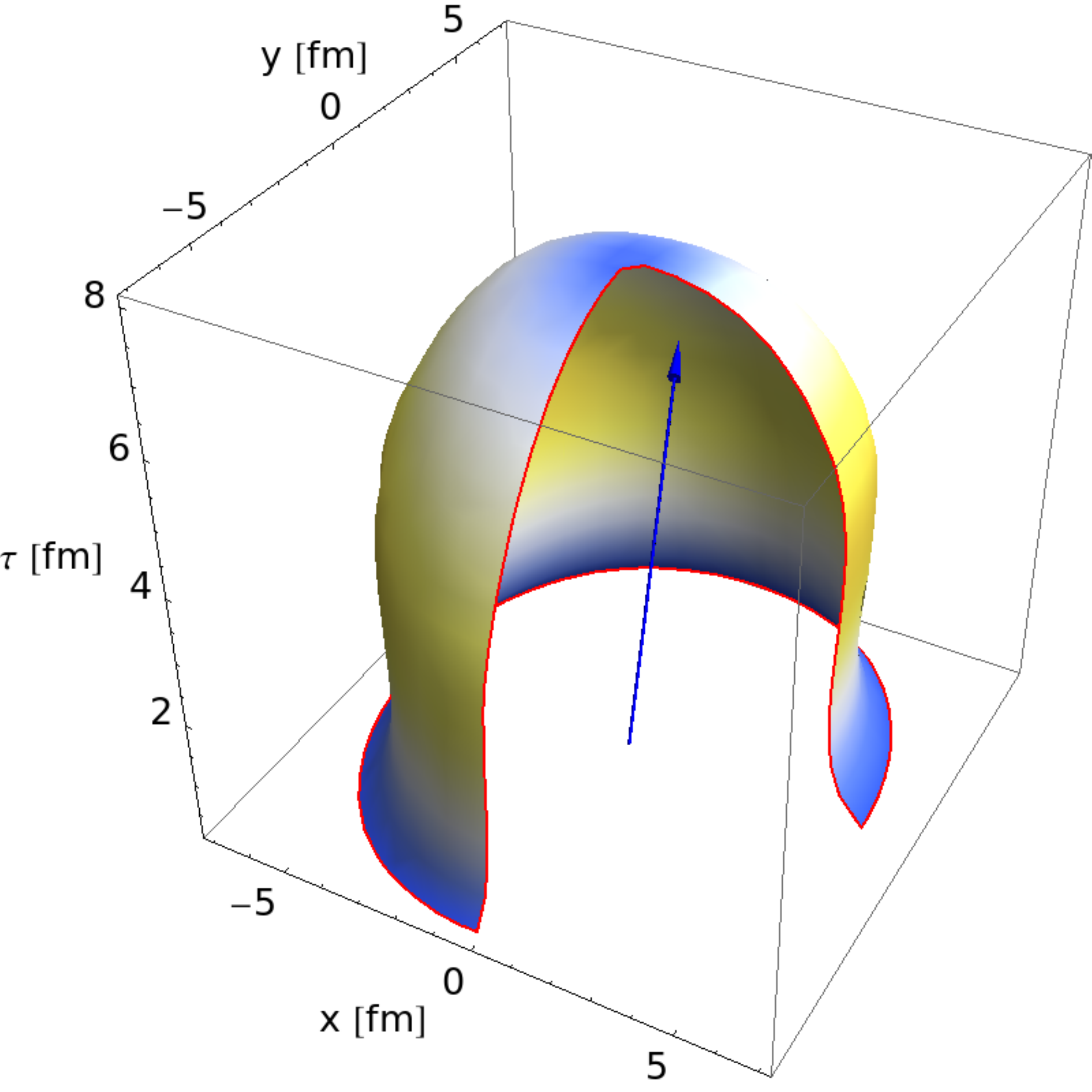}
\end{center} 
 \caption{Isothermal freeze-out surface in Milne coordinates ($\varsigma=0$).}
 \label{fig:fo3d}
\end{figure}
%
When initialized with proper initial conditions the perfect fluid dynamics determines  the evolution of the temperature, flow velocity, and the chemical potentials (if charge conserving theory is considered) in the entire forward light-cone of the Minkowski space-time ${\cal M}$, whose points satisfy the requirement $\tau \ge \tau_{\rm i}$. Usually, due to specific shape of the isothermal freeze-out hypersurface $\Sigma$ (see Figs.~\ref{fig:evo} and \ref{fig:fo3d}), it is convenient to parametrize the ambient space-time ${\cal M}$ with three angles, say $\theta,\zeta$ and $\phi$, and the distance  $\rho$ from the coordinate system's origin $(\tau=\tau_{\rm i},x=0,y=0,\varsigma=0)$~\footnote{Note from Figs.~\ref{fig:evo} and \ref{fig:fo3d} that $\tau$ is not always a function of remaining space-time coordinates.}. The resulting parametrization reads \cite{Bozek:2009ty, Ryblewski:2013jsa}
%
\begin{figure}[t]
\begin{center}
\includegraphics[angle=0,width=0.49 \textwidth]{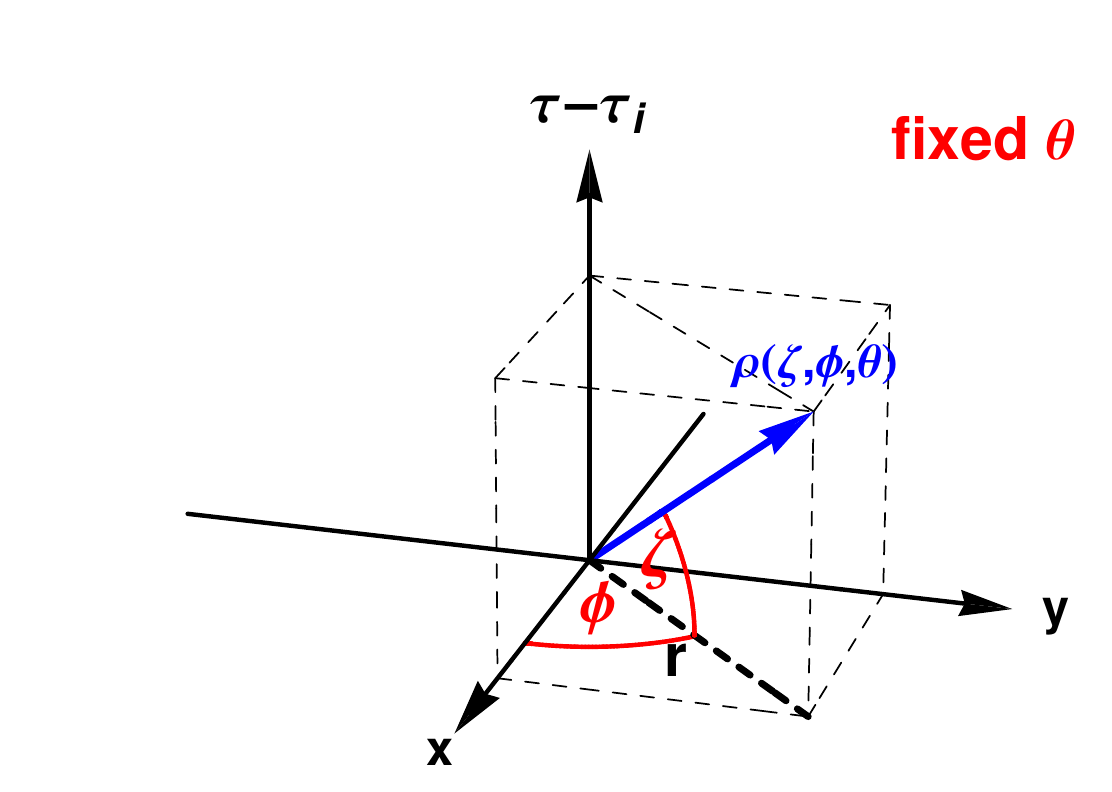}
\includegraphics[angle=0,width=0.49 \textwidth]{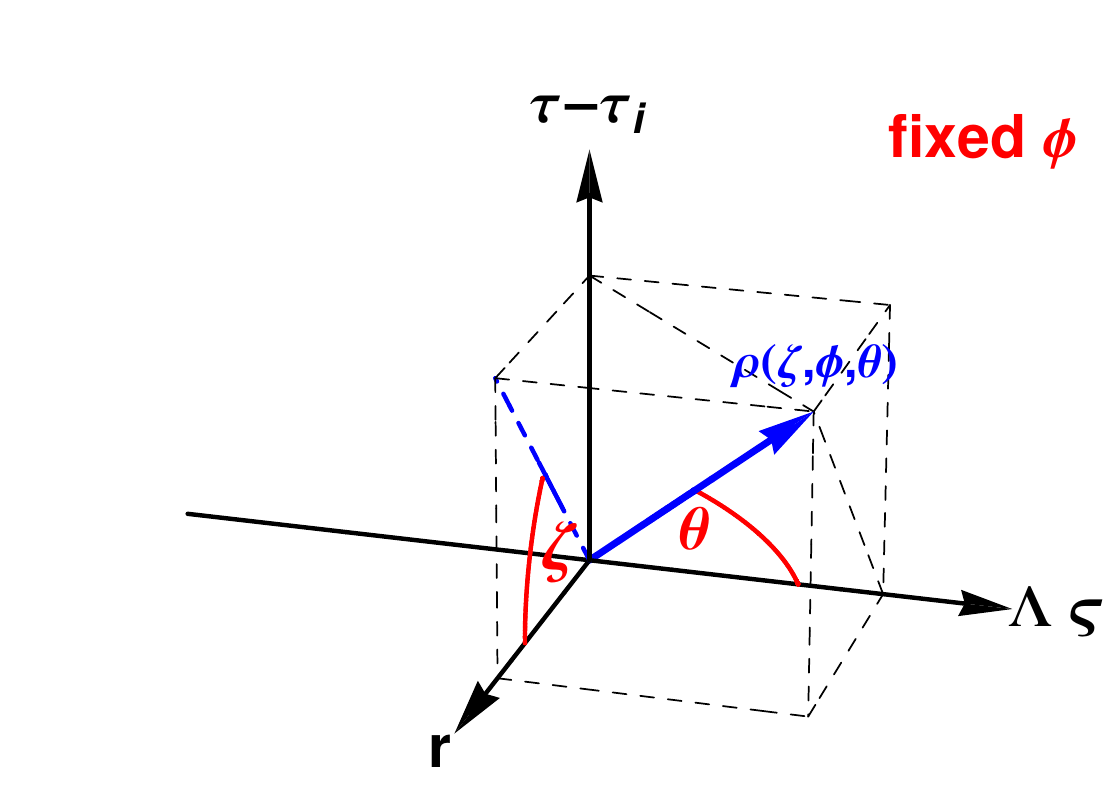}
\end{center} 
 \caption{Space-time parametrization with angles $\theta,\zeta$ and $\phi$, and the distance  $\rho$.}
 \label{fig:parametr}
\end{figure}
%
%
\begin{eqnarray} 
x^0=t(\rho,\theta,\zeta,\phi)  &=& \tau \cosh \varsigma , \\
x^1=x(\rho,\theta,\zeta,\phi)  &=& r \sin\theta\cos\phi   , \\
x^2=y(\rho,\theta,\zeta,\phi)  &=& r \sin\theta \sin\phi  ,\\
x^3=z(\rho,\theta,\zeta,\phi)  &=& \tau \sinh \varsigma,
\label{freezepar}
\end{eqnarray}
with 
\begin{eqnarray}
\Lambda\,\varsigma  &=&  \rho \cos\theta , \\
 r &=&  \rho \cos\zeta,
\label{freezepar2}
\end{eqnarray}
so that $\tau -\tau_{\rm i} =  \rho \sin\theta \sin\zeta$, and 
\begin{eqnarray}
0 \leq  \zeta    \leq \pi/2, \qquad
0 \leq \phi   < 2 \pi,  \quad {\rm and} \quad 
0 \leq \theta  \leq \pi;
\label{angleslim}
\end{eqnarray} 
see Fig.~{\ref{fig:parametr}}. The isothermal freeze-out condition defines the  hypersurface $\Sigma$ embedded in ${\cal M}$ in the following way 
\begin{eqnarray}
\Sigma(\theta,\zeta,\phi) = \{x \in {\cal M}: T(\rho,\theta,\zeta,\phi) = T_{\rm freeze} \},
\label{constT}
\end{eqnarray}
so that one has implicitly $\rho=\rho(\theta,\zeta,\phi)$. One should note here that, while at $\Sigma$, by definition, the temperature is fixed, the flow four-velocity is not,  $u^{\mu}=u^{\mu}(\theta,\zeta,\phi)$. The infinitesimal element of the hypersurface $\Sigma$ is defined by the covariant four-vector \cite{Misner:1974qy} 
\begin{equation}
\dS_\mu= \varepsilon_{\mu \alpha \beta \gamma}
\frac{\partial x^\alpha}{\partial \zeta} \frac{\partial x^\beta}{\partial \phi} \frac{\partial x^\gamma}{\partial \theta }
d\zeta d\phi d\theta,
\label{d3Sigma}
\end{equation}
where $\varepsilon_{\mu \alpha \beta \gamma}$ is the totally antisymmetric tensor in four dimensions with $\varepsilon_{0123} = +1$.
%
\section{Cooper-Frye formalism}
\label{sec:6}
\sectionmark{Cooper-Frye formalism}
%
The change from fluid elements to hadrons at the switching hypersurface $\Sigma(x)$ is usually performed with the use of kinetic theory concepts. In the transport theory framework the flux of particle species $k$ is expressed with the formula (see Eqs.~(\ref{mom1}) and (\ref{ident1})) \cite{Rischke1999}
\begin{equation}
N_k^\mu(x) = \int\frac{\dP}{E_p} p^\mu f_k(x,p), 
\label{partflux}
\end{equation}
where  $p^\mu=(E_p, \textbf{p}, p_z)$ is the four-momentum of the (on-shell) particle with mass $m_k$, and $f_k(x,p)$ is its phase-space distribution function. The number of world-lines of particles of species $k$ crossing the infinitesimal element $\dS$ of the surface $\Sigma$ is then calculated from the expression
\begin{equation}
 N^k_{\Sigma} \equiv \dS_\mu N_k^\mu = \int\frac{\dP}{E_p} \dS_\mu p^\mu f_k(x,p).
\label{partnum}
\end{equation}
The invariant momentum spectrum of particles produced at the surface element $\dS$ is
\begin{equation}
E_p\frac{dN^k_{\Sigma}}{\dP} = \dS_\mu  p^\mu f_k(x, p). 
\label{momspec}
\end{equation}
Therefore, the invariant momentum distribution of hadrons emitted on the entire freeze-out hypersurface $\Sigma$ is given by the integral  
\begin{equation}
E_p\frac{dN^k}{\dP} =  \int_\Sigma   \dS_\mu  p^\mu f_k(x, p).
\label{cfform}
\end{equation}
Equation~(\ref{cfform}) is commonly known as the Cooper-Frye formula \cite{Cooper:1974mv}.

It is usually assumed that just before decoupling from the fluid, \emph{i.e.}, before the particles cross the switching surface $\Sigma$, they are in local thermal and chemical equilibrium such that their phase-space distributions are described by the equilibrium ones. In such a case one assumes that produced hadrons follow  either  Fermi--Dirac ($a=-1$)  or  Bose--Einstein ($a=+1$)  distributions,
\begin{equation}
 f_k(x, p)= f_k\left(p_\mu u^\mu(x), T(x), \tilde{\mu}_k\right) = \frac{g_k}{(2\pi)^3}\left\{ \exp\left[\frac{p_\mu  u^\mu(x)  -   \tilde{\mu}_k(x) }{T(x)}\right] -a \right\}^{-1}\!\!\!,
\label{eqdistr}
\end{equation}
where the factor $g_k = 2s_k+1$ takes into account the spin degeneracy of hadron species $k$.
The chemical potential $\tilde{\mu}_k$ is given by the linear combination of the chemical potentials $\mu_i$ (see Sec.~\ref{sec:2a}) and the respective charges of the hadron species $k$ 
\begin{equation}
  \tilde{\mu}_k = \sum \limits_i Q_i^k \mu_i.
  \label{chempot}
\end{equation} 
The particle's four-momentum, as measured by the experiment, is usually parametrized in the following way
\begin{equation}
p^{\mu} = \left( m_T \cosh {\rm y}_p, p_T \cos \phi_p, p_T \sin \phi_p, m_T \sinh  {\rm y}_p\right),
\label{particlemom}
\end{equation}
where $p_T=\sqrt{p_x^2+p_y^2}$ is the transverse momentum, $m_T=\sqrt{p_T^2+m_k^2}$ is the transverse mass, $y_p= \ln\left[ (E_p+p_z)/(E_p-p_z)\right]/2$ is the longitudinal rapidity and $\phi_p = \tan^{-1} \left(p_y/p_x\right)$ is the momentum azimuthal angle in the  plane transverse to the beam axis. 

With definitions introduced above the integration measure $ \dS_\mu p^\mu $ in Eq.~(\ref{cfform}) takes the form
\begin{eqnarray}
  \dS_\mu p^\mu 	
&=& \frac{\sin\theta \tau  \rho^2  }{\Lambda} 
\Biggl[ \Biggr.  
\frac{\partial\rho}{\partial\zeta} \cos\zeta 
\left(
p_T \sin\zeta \cos\left(\phi_p - \phi\right)- m_T \cos\zeta \cosh\left({\rm y}_p - \varsigma \right) 
\right) 
\nonumber\\
&+& \!\! \cos\zeta \sin\theta \left(\rho \sin\theta - \frac{\partial \rho}{\partial\theta} \cos\theta \right)
\nonumber\\
&\times&
\left(
 p_T \cos\zeta \cos\left(\phi_p - \phi\right) + m_T \sin\zeta \cosh\left({\rm y}_p- \varsigma \right)
\right) 
\nonumber\\
&+& \!\! \cos\zeta \sin\theta \left(\rho \cos\theta + \frac{\partial \rho}{\partial\theta}   \sin\theta \right) \frac{\Lambda}{\tau} m_T  \sinh\left({\rm y}_p - \varsigma \right) 
\nonumber\\
&-& \!\! \frac{\partial \rho}{\partial\phi}\,\, p_T\, \sin(\phi_p - \phi) 
\Biggl. \Biggr]  
d \zeta d\phi d\theta \equiv   h_k(\zeta,\phi,\theta, p_T, \phi_p, {\rm y}_p)d \zeta d\phi d\theta,
\label{intmeas}
\end{eqnarray}
and the (Lorentz-boosted) energy is 
\begin{equation}
p_\mu u^\mu = u_0\,   m_T \cosh ({\rm y}_p-{\rm y}_u) -   p_T   u_T \cos(\phi_p-\phi_u),
\label{pu}
\end{equation}
where we introduced yet another variable $\phi_u$ such that $u_x = u_T \cos \phi_u$ and $u_y = u_T \sin \phi_u$.
One should note that in cases where some symmetries are present in the system the formulas (\ref{intmeas})-(\ref{pu}) may be respectively simplified \cite{Florkowski:2010zz,Chojnacki:2007rq}.
Using expressions (\ref{intmeas})-(\ref{pu}) in the Cooper-Frye formula (\ref{cfform}) one obtains a six-dimensional particle distribution, which can be used directly to generate particles (both, stable hadrons and unstable resonances) on $\Sigma$
\begin{eqnarray}
 \frac{d^6N^k}{p_T d p_T d\phi_p d{\rm y_p}   d\zeta d \phi d \theta} &&= \frac{g_k}{(2 \pi)^3} h_k (\zeta,\phi,\theta, p_T, \phi_p, {\rm y}_p) f_k(\zeta,\phi,\theta, p_T, \phi_p, {\rm y}_p) \nonumber \\
 &&\equiv {\cal F}_k (\zeta,\phi,\theta, p_T, \phi_p, {\rm y}_p).
\label{cfformTH}
\end{eqnarray}

The Cooper-Frye formula, Eq.~(\ref{cfform}), is nowadays commonly used in the fluid dynamical simulations of the  heavy-ion collisions to describe hadron production on the freeze-out hypersurface. There are, however, well known limitations of this prescription. An immediate problem with Eq.~(\ref{cfform}) arises if the freeze-out hypersurface contains both time-like and space-like parts. In particular, if the freeze-out element is time-like, so that associated normal vector is space-like, for certain directions of the momentum $p^\mu$ the invariant measure $\dS_\mu p^\mu$ may be negative. Thus, the particle number generated in this region will become ill-defined (the Cooper-Frye formula would in such a case describe the back-flow of particles into the fluid) \cite{Rischke1999}. One may show that at high energies the negative emission is usually negligible \cite{Chojnacki:2011hb}, so that it may be safely removed by introducing the step function $\Theta(\dS_\mu p^\mu)$ on the right-hand side of Eq.~(\ref{cfform}) \cite{Rischke1999}. Another issue connected with Eq.~(\ref{cfform}) is its insensitivity to the fact that particles with large momenta are in general more probable to leave the fluid more easily than the soft ones \cite{Kolb:2003dz}.
%
\section{Hadron abundances}
\label{sec:7}
\sectionmark{Hadron abundances}
%
For the system in local thermal equilibrium the flux $N_k^\mu$ of particle species $k$ from   the fluid cell is proportional to its four-velocity, see Eq.~(\ref{pflux}). In this case, using formula (\ref{partnum}), the total number of particles emitted on the entire hypersurface may be expressed as follows 
\begin{equation}
 N_k  \equiv \int_\Sigma \dS_\mu N_k^\mu = \int_\Sigma \dS_\mu u^\mu(x) \n_k\left(T(x), \tilde{\mu}_k(x)\right).
\label{avpart}
\end{equation}
One should stress here that the particle density $\n_k$ in Eq.~(\ref{avpart}) is expressed solely through the local temperature $T(x)$ and chemical potentials $\tilde{\mu}_k(x)$. It straightforward to see that, if $T$ and  $\tilde{\mu}_k$ are constant along $\Sigma$, the integral of the flow pattern on the freeze-out manifold factorizes in Eq.~(\ref{avpart}), giving the so called effective comoving volume $V_{\rm eff}\equiv\int_\Sigma \dS_\mu u^\mu(x)$. When one considers ratios of particle multiplicities of different species, say $a$ and $b$, the factor $V_{\rm eff}$ cancels out completely in the ratios \cite{Heinz:1998st,Cleymans:1998yf} giving
\begin{equation}
 \frac{N_a}{N_b}  = \frac{\n_a(T, \tilde{\mu}_k)}{\n_b(T, \tilde{\mu}_k)}.
\label{ratios}
\end{equation}
Arguments presented above gave rise to the wide variety of analyses under the common name of \emph{thermal} or \emph{statistical} models \cite{BraunMunzinger:2001ip,Florkowski:2001fp,Rafelski:2001hp,Baran:2003nm,Turko:2007ri}. They focus mainly on the extraction of thermodynamic properties of the matter at the chemical freeze-out based on thermal analysis of the   multiplicities of the experimentally measured particles and ratios thereof. Using the grand canonical version of the thermal approach the fits usually yield the chemical freeze-out temperature of the order of the quark-hadron phase transition obtained from the lattice QCD  calculations
\begin{equation}
T_{\rm chem} \sim T_{\rm c} \sim 170 \,\,\,{\rm MeV} ,
\label{Tchem}
\end{equation}
which suggest possible relation between hadronization process and chemical equilibration \cite{BraunMunzinger:2003zz} \footnote{Note that the new lattice QCD simulations suggest somewhat lower values the critical temperature, $T_{\rm c} \sim 155 \,\,\,{\rm MeV}$ \cite{Borsanyi:2010cj}.}.
 
Equation (\ref{ratios}) requires a few important remarks: (i) the integrals quoted above are performed over the full momentum space, which means that the reliable analysis would require  the $4\pi$ acceptance for identified particles, (ii) at low energies and forward rapidities the freeze-out conditions are usually quite different from the baryon-free midrapidity region, suggesting that thermal analyses based on Eq.~(\ref{ratios}) yield, in this case, only approximate (averaged over the entire hypersurface) values of thermal parameters at   freeze-out. Nevertheless, keeping in mind its simplicity, the precision of the thermal approach is quite remarkable.
 
\section{Decays of resonances}
\label{sec:9}
\sectionmark{Decays of resonances}
%
When discussing the thermal/statistical approach we neglected an important aspect of the modeling connected with the role of resonances. It is usually assumed that the particles created at   freeze-out form an ideal non-interacting gas of hadrons, which includes, in principle, entire mass spectrum of unstable resonances. In reality, the latter subsequently decay populating the spectrum of stable hadrons, which is then observed in the detector. Although the Boltzmann factor tends to suppress heavy states, one should keep in mind that, according to the Hagedorn hypothesis \cite{Hagedorn:1965st}, their mass spectrum increases exponentially. While at low temperatures the role of the resonance feed-down is diminished, at the temperatures of the order of $T_{\rm chem}$ it is quite significant. In fact, the successful description of the available data on the hadronic abundances within the statistical approach, as quoted in Sec.~\ref{sec:7}, was possible largely due to inclusion of the full mass spectrum of the hadronic resonances \cite{Amsler:2008zzb} \footnote{In practice, all resonances from the Particle Data Group tables \cite{Amsler:2008zzb}, whose properties are known well enough, were included.}. 

As it was shown within the  so called  Cracow model \cite{Broniowski:2001uk} (see next Section), which includes hydrodynamic-like expansion of the system, the role of the resonance feed-down turns out to be equally important for   description of the momentum spectra of stable hadrons as the flow itself. This is mainly due to the fact that   decays of heavy resonances populate mainly the soft region of the stable hadron transverse-momentum spectra leading to their steeper slopes. Effectively, the observed inverse slope parameter, usually interpreted as the thermal freeze-out temperature, is much smaller than the chemical freeze-out temperature inferred from the analysis of the hadronic ratios
\begin{equation}
T_{\rm therm} \sim 130 \,\,\,{\rm MeV};
\label{Ttherm}
\end{equation}
compare Eq.~(\ref{Tchem}). This observation gave further support to the single-freeze-out model, and explained apparent mismatch between the two freeze-out temperatures.
%
\section{Hydro-inspired parameterizations of freeze-out}
\label{sec:10}
\sectionmark{Hydro-inspired parameterizations}
%
According to Eq.~(\ref{eqdistr}) the spectrum of particles produced in a single fluid cell is thermal. However, even if the thermal parameters $T$ and $\tilde{\mu}$ are constant along the $\Sigma$, the total momentum spectrum, as calculated with Eq.~(\ref{cfform}), includes contributions from different fluid cells, each boosted with a different velocity $u^\mu(x)$. Therefore, the resulting total spectrum is modified due to the combination of \emph{redshift} and \emph{blueshift} effects \cite{Broniowski:2001we}. These effects are observed in the experiment in the form of the characteristic concave shape of the transverse momentum spectrum. In view of these arguments, the realistic description of the momentum spectra of stable hadrons must include effects of some kind of collective evolution reflected in the finite flow of the matter at   freeze-out. 

The most natural way to include flow at the freeze-out is to perform full fluid dynamical simulations along the lines presented in Sec.~\ref{sec:2a}. Unfortunately, numerical solution of fluid dynamical equations of motion is rather complicated and computationally intensive. Moreover, fluid evolution requires at least \emph{some} knowledge of the initial conditions, which are rather poorly known, and thus always bias final results. In order to avoid such problems, one may follow a different strategy, which results in the so called \emph{hydro-inspired} models. Within these models the conditions at the freeze-out hypersurface  $\left(T(x), u^\mu(x), \mu_i(x)\right)$, as well as the shape of the freeze-out surface $\Sigma(x)$, are simply assumed, or \emph{inspired}, by the full numerical fluid dynamical simulations. Among these models the most successful ones are the Blast-wave \cite{Kisiel:2006is,Schnedermann:1993ws} and Cracow  \cite{Broniowski:2001we} models.

In particular, within the Cracow model \cite{Broniowski:2001we} (as well as in Blast-wave model) it is assumed that the freeze-out takes place at the boost-invariant  and cylindrically symmetric (in the transverse $x-y$  plane). The hypersurface $\Sigma$ is defined by the requirement that the particles freeze-out at the surface of constant proper-time (${\tilde{\tau}}$)
\begin{equation}
{\tilde{\tau}}^2 = x^\mu x_\mu = t^2 - x^2 - y^2 - z^2 =
\tau^2 -r^2 = {\tilde{\tau}}_{\rm freeze}^2 = \hbox{const.}
\label{hubtau}
\end{equation}
According to Eq.~(\ref{hubtau}) the particles decouple starting from the center of the \emph{fire-cylinder} towards its edge, so that $0\le r \le r_{\rm max}$. At the same time the fluid four-velocity is assumed to have the Hubble-like form \cite{Chojnacki:2004ec}
\begin{equation}
u^\mu = \gamma (1, {\bf v}_T, v_z) = {x^\mu \over \tilde{\tau}_{\rm freeze}},
\label{hubumu}
\end{equation}
which leads to 
\begin{eqnarray}
  p_\mu u^\mu		&=&   \frac{\sqrt{\tilde{\tau}_{\rm freeze}^2 + r^2}}{\tilde{\tau}_{\rm freeze}}\,\, m_T \cosh(y_p-\varsigma) -   p_T \frac{r}{\tilde{\tau}_{\rm freeze}} \cos(\phi_p-\phi),
\label{Cracowpu}
\end{eqnarray}
compare Eq.~(\ref{pu}).
Conditions  (\ref{hubtau}) and (\ref{hubumu}) imply that the freeze-out hypersurface element  is proportional to four-velocity 
\begin{eqnarray}
\dS   	&=&  u^\mu \tilde{\tau}_{\rm freeze}\, r dr\, d\varsigma\, d\phi.
\label{CracowdS}
\end{eqnarray}
Within the Cracow model, and with the inclusion of all known hadronic resonances, it was possible to describe, both, the particle ratios  and spectra of particles \cite{Broniowski:2001uk}, within the single framework. 

Due to the fast increase of the computer power during the last decade the full numerical solutions of fluid dynamical evolution equations, including the dissipative effects became easily achievable. As a result the importance of hydro-inspired models was largely reduced, making them useful tools for the estimate of flow and thermal properties at the freeze-out mainly in the analyses of the experimental data; although see \cite{Begun:2013nga}.
%
\section{Monte-Carlo statistical hadronization}
\label{sec:11}
\sectionmark{Statistical hadronization with \THERMINATOR }
%
Careful analysis of the experimental data on flow and correlations measured at RHIC and the LHC energies has shown that the realistic description of the dynamics of heavy-ion collisions requires, an experimental-wise, \emph{event-by-event} simulations of such reactions. Due to event-by-event initial state fluctuations (of different kind), such a modeling usually involves running fluid dynamical evolution separately for each event. As a result, in each event the flow patterns at $\Sigma$, as well as the shape of the $\Sigma$, does not exhibit any symmetries. Hence, in general, the calculation of the particle spectra using Cooper-Frye formula, Eq.~(\ref{cfform}), has to be done entirely numerically. 
Moreover, precision studies usually require various experimental cuts and feed-down corrections to be applied to reproduce the data correctly. Thus, it is crucial to have the access to the full information on phase-space properties of the produced hadrons \footnote{The access to the entire phase-space properties of the created particle ensemble is of great importance (note that the experimental analysis may access only the four-momentum properties of the particles). As a result, one may, for instance, relate the experimental-wise calculated HBT radii of the system with its \emph{actual} space-time size in simulations.}. In view of these arguments, the development of Monte-Carlo generators for simulation of physical events became necessary. One of the first numerical open-source codes devoted to this task was \THERMINATOR (THERMal heavy-IoN generATOR) \cite{Kisiel:2005hn} (for its new, extended version  -- \THERMINATORTWO -- see \cite{Chojnacki:2011hb} \footnote{The \THERMINATORTWO code \cite{Chojnacki:2011hb} was also supplemented with another (separate) code, \texttt{FEMTO-\THERMINATOR},
which is provided to carry out the analysis of the pion--pion femtoscopic correlations.}). 

The \THERMINATOR's main functionality is to perform hadronization in relativistic heavy-ion collisions using concepts of the statistical approach and the single-freeze-out model. In its latest version \cite{Chojnacki:2011hb}, the code performs event-by-event generation of an ensemble of particles (an equivalent of the physical event) given any freeze-out conditions (shape of the hypersurface, flow, and thermodynamic parameters) typically created in fluid dynamic models  using the Cooper-Frye formula~(\ref{cfform}). Within the code one may choose, either  one of the predefined hydro-inspired freeze-out parameterizations (such as Blast-wave, or Cracow; see Sec.~\ref{sec:10})  or  an output from any realistic fluid dynamical simulations. Since nowadays the fluid dynamical modeling became the common standard in the field of heavy-ion collisions (at least with event-averaged initial conditions), in what follows we   focus mainly on the latter case.

In the case of using directly the fluid dynamics simulations, at first the freeze-out has to be extracted from the fluid evolution, which is out of the scope of \THERMINATOR  code. In particular, in the case of perfect fluid dynamics, one has to supply the distance $\rho(\zeta,\phi,\theta)$, the components of the flow velocity $u_x(\zeta,\phi,\theta)$, $u_y(\zeta,\phi,\theta)$ and $y_u(\zeta,\phi,\theta)$, temperature $T(\zeta,\phi,\theta)$, and chemical potentials $\mu_i(\zeta,\phi,\theta)$ (although  the values of chemical potentials are usually neglected in the hydrodynamic stage) \footnote{In the case of isothermal freeze-out the temperature $T(\zeta,\phi,\theta)=T_{\rm freeze}$ is  constant. Chemical potentials are usually included at the freeze-out, based on the results from the thermal model fits $\mu_i(\zeta,\phi,\theta)$.}. For the perfect fluid case, it is assumed that on the hypersurface $\Sigma(x)$ the system is in local thermal and chemical equilibrium, which means that the phase-space distributions of the particles have the forms given by Eq.~(\ref{eqdistr}) \footnote{In the case of using output from the dissipative fluid dynamics one should correct the distribution function (\ref{eqdistr}) for the non-equilibrium effects as well as supply the \THERMINATORTWO  code with the respective dissipative quantities at $\Sigma$.}. 

 The \THERMINATOR's code is written in an object-oriented C++ programing language and conforms to the CERN's \texttt{ROOT} framework standards \cite{Brun:1997pa}. Once the proper input is provided, the generation of hadrons (stable ones and resonances) is done using straightforward Monte Carlo method according to the Cooper-Frye formula (\ref{cfformTH}). The detailed description of the \THERMINATOR's theoretical background, code structure, functionalities, as well as short introduction to its usage may be found in the original papers \cite{Kisiel:2005hn,Chojnacki:2011hb}, and   on the project's website \cite{THERMINATORweb}. Herein we will just briefly review its main aspects.  
 
Each evaluation of the code is performed in two stages. The first (preliminary) stage, which is performed once per set of parameters, for each particle species, and whose results are recorded  and used for all subsequently generated events, involves:
\begin{itemize}
\item Calculation of the global maximum ${\cal F}^k_{\rm max}$ of the right-hand side of Eq.~(\ref{cfformTH}).

\item Calculation of the average multiplicity $\bar{N}_k$ by integrating  Eq.~(\ref{cfformTH}) over the entire phase-space (\emph{i.e}. over $\zeta,\phi,\theta, p_T, \phi_p$ and ${\rm y}_p$).  
\end{itemize}

The second (main) stage consists of generation of (primordial) particle ensemble and performing decays of unstable resonances, which proceeds on event-by-event basis. In each event:
 \begin{itemize}
 \item It is assumed that the generated ensemble of particles corresponds to the grand canonical ensemble. Hence, the number $N_k$ of particles of species $k$ in the event is generated randomly according to the probability expressed by the Poisson distribution 
\begin{equation}
 P(N_k)=\frac{ \left(\bar{N}_k\right)^{N_k}}{N_k!}\exp(-\bar{N}_k)
\label{Poisson}
\end{equation}
For each particle species $k$ number  $N_k$ of particles is generated according to the von-Neumann acceptance/rejection procedure: space-time point $(\zeta,\phi,\theta)$ at $\Sigma$, momentum components of the particle ($p_T, \phi_p$ and ${\rm y}_p$), and a test variable ${\cal F}^k_{\rm test}$ in the range $\langle 0, {\cal F}^k_{\rm max}\rangle$, are generated randomly. The particle is accepted if ${\cal F}^k_{\rm test}<{\cal F}^k(\zeta,\phi,\theta,p_T, \phi_p, {\rm y}_p)$, otherwise it is rejected. The generation of particles goes over all species (stable ones and resonances), which are listed in the Particle Data Group \cite{Amsler:2008zzb} tables, and whose properties are known well enough. For that purpose the \SHARE particle database is used \cite{Torrieri:2004zz}. 

\item Once the ensemble of primordial particles is generated the code performs decays of unstable resonances, which, in general, may proceed in cascades. Each resonance evolves along the classical trajectory starting from its
initial position $x^\mu_{\rm origin}$ according to its momentum
\begin{equation}
x^\mu_{\rm decay} = x^\mu_{\rm origin} + {p^\mu \over m_k} \Delta \tau.
\label{decpt}
\end{equation}
and decays after its lifetime $\Delta\tau$, which is randomly generated with the probability density $\Gamma_k \exp(-\Gamma_k \Delta\tau)$, where $\Gamma_k$ is the width of the particle of species $k$. The particular decay channel is selected randomly with the probability corresponding to its branching ratio. The decays of sub-threshold type are not allowed. The decays, which are of two-particle or three-particle type, follow simple kinematic formulas \cite{Kisiel:2005hn}, and are treated on equal footing. All required data on the decays follows from the \SHARE~ particle decays database \cite{Torrieri:2004zz}.

\item Once all particles in the event decayed, the calculation is completed.
  
\end{itemize}

\begin{figure}[t]
\begin{center}
 \includegraphics[angle=0,width=0.49 \textwidth]{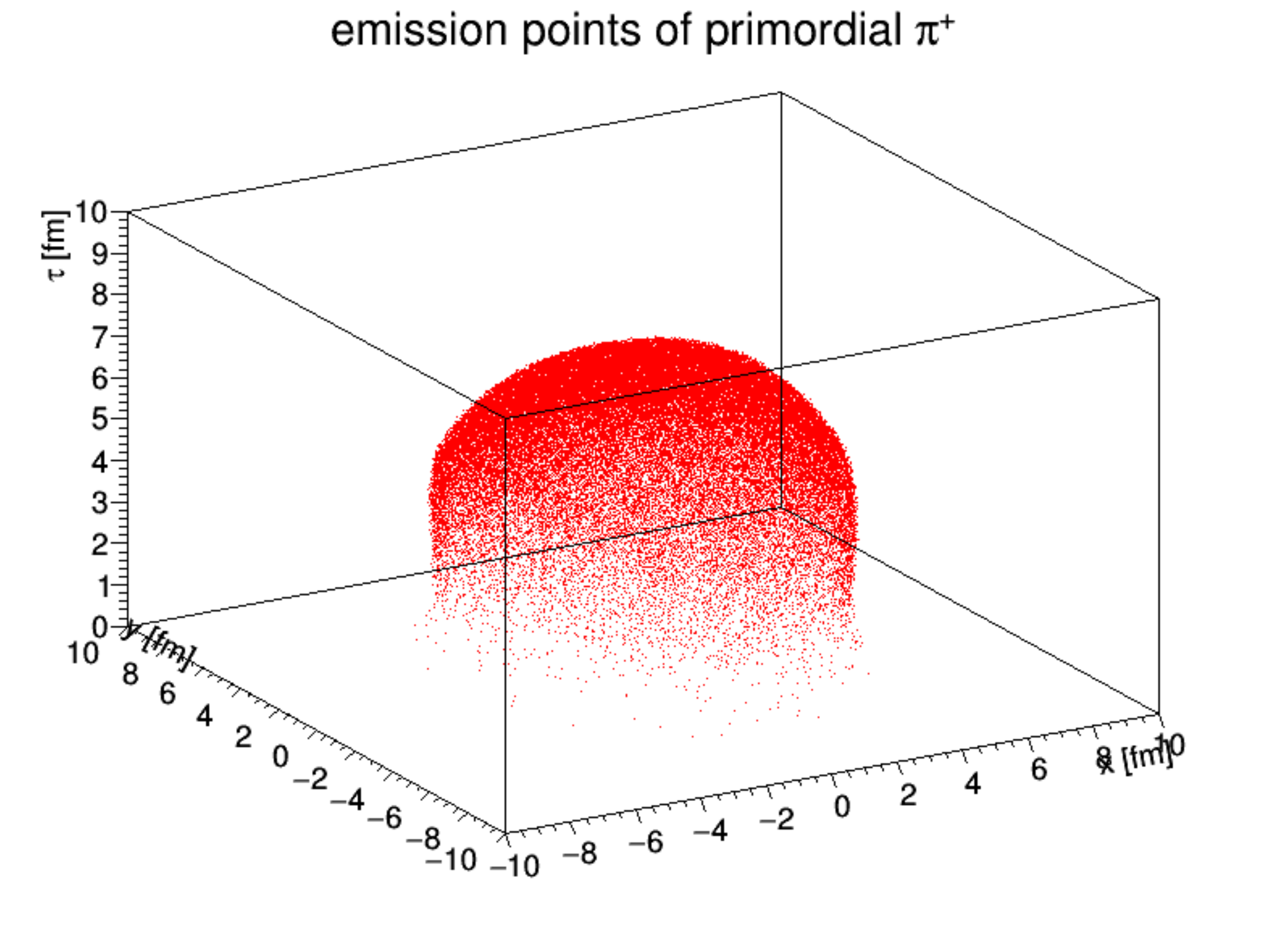} \includegraphics[angle=0,width=0.49 \textwidth]{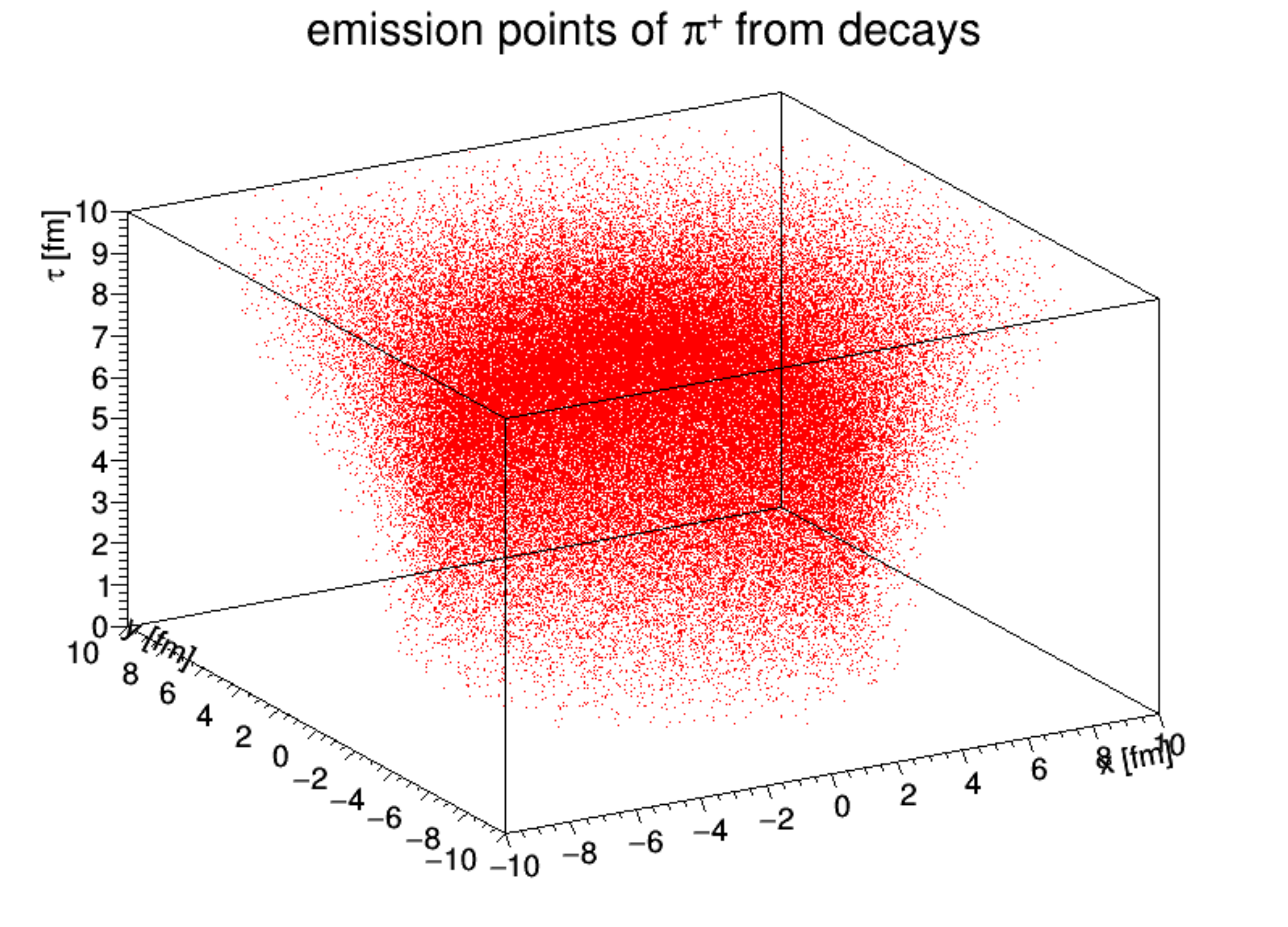} 
\end{center}
 \caption{Emission points of primordial $\pi^{+}$ (left panel), and $\pi^{+}$ from decays (right panel) in the $x-y-\tau$ plane in Au-Au collisions at $\sqrt{s_{\rm NN}}= 200~{\rm GeV}$ and the impact parameter of $7.16$ fm.  }
\label{fig:emiss}        
\end{figure}

The exemplar emission points in space-time obtained with the initial tilted source and \THERMINATOR simulations are presented in the Fig.~\ref{fig:emiss}. In the next Section, based on the data on the emitted particles, we will calculate some physical observables.
%
%
%
\section{Performing analysis with \THERMINATORTWO}
\label{sec:12}
\sectionmark{}
%
Due to the detailed record on properties of all produced particles, including their space-time coordinates $(x^\mu, p^\mu)$  and their decay chains, the \THERMINATOR~code becomes a versatile tool, allowing for calculation of various observables. Having completed generation of events various experimental observables may be calculated, either, by using the figure macros supplied with the code, or, by preparing user macros. In this Section we will present some of its capabilities. 
%
\begin{figure}[t]
\begin{center}
 \includegraphics[angle=0,width=0.65 \textwidth]{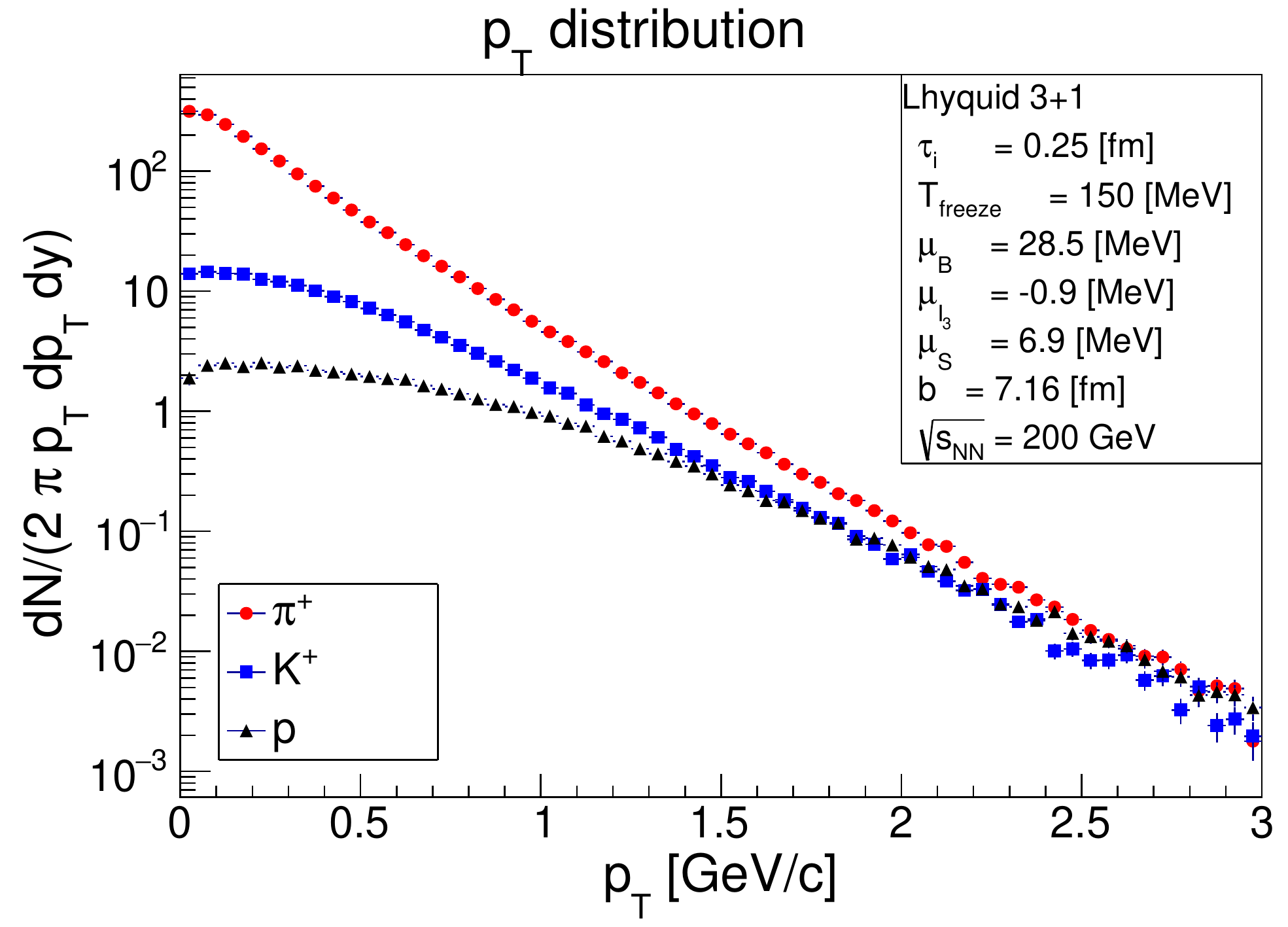}  
 \includegraphics[angle=0,width=0.65 \textwidth]{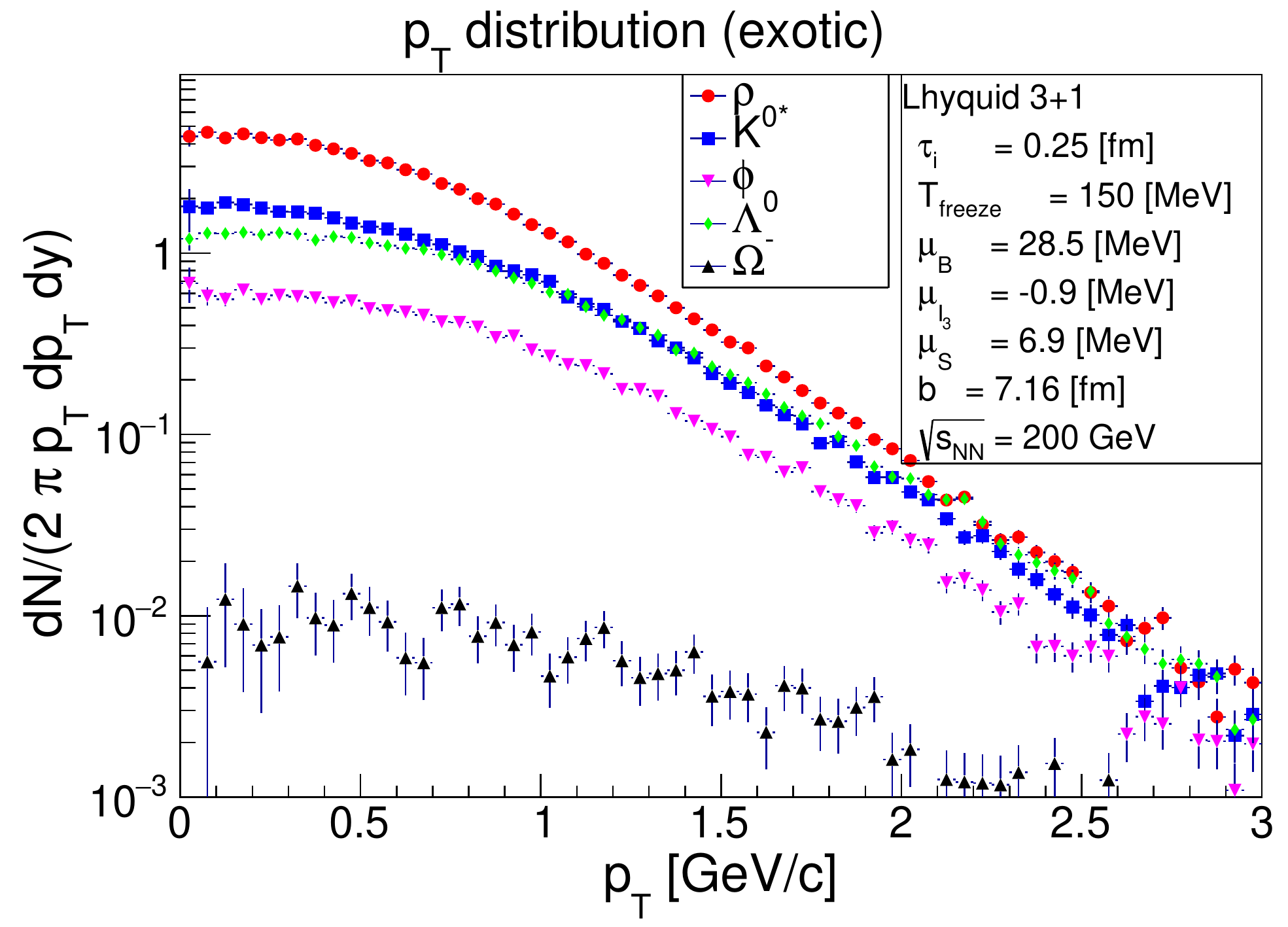} 
\end{center}
 \caption{Transverse-momentum spectra of $\pi^+$, $K^+$, and protons (top panel) and $\rho$ mesons, $K^*_0$
mesons, $\phi$ mesons, $\Lambda^0$ barions,
and $\Omega^-$ barions (bottom panel) for Au-Au collisions at $\sqrt{s_{\rm NN}}= 200~{\rm GeV}$ and the impact parameter of $7.16$ fm (protons from the weak decays of $\Lambda$'s are excluded). The statistics of 3000 events was used. The errors are statistical only. }
\label{fig:spec}        
\end{figure} 
%
\subsection{Single-particle spectra}
\label{ssec:121}
\begin{figure}[t]
\begin{center}
 \includegraphics[angle=0,width=0.65 \textwidth]{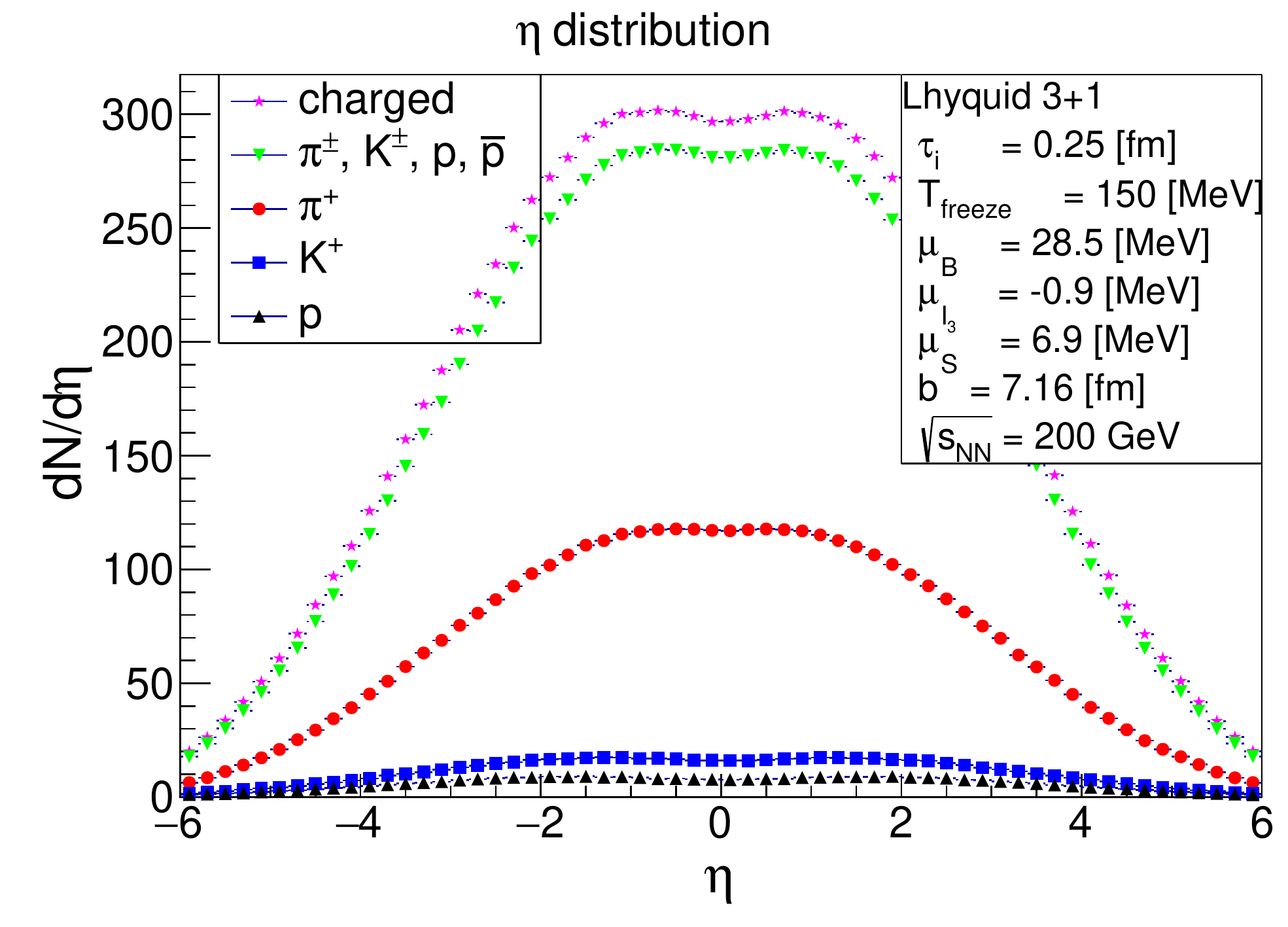} 
\end{center}
 \caption{Pseudorapidity spectra  of $\pi^{+}$, $K^{+}$, and protons shown separately, and summed together with their antiparticles, as well as of all charged particles,   for Au-Au collisions at $\sqrt{s_{\rm NN}}= 200~{\rm GeV}$ and the impact parameter of $7.16$ fm. The statistics of 3000 events was used. The errors are statistical only.}
\label{fig:specpseu}        
\end{figure}

\sectionmark{hs}
%
One of the most straightforward observables to calculate is the single-particle spectrum of certain particle species $k$. The most copiously produced ones are the lightest mesons (pions and kaons) and baryons (protons). They form more than $90\%$ of the total charged particles.
In the Fig.~\ref{fig:spec} we present the transverse-momentum spectra of $\pi^+$, $K^+$, and protons (top panel), as well as $\rho$ mesons, $K^*_0$ mesons, $\phi$ mesons, $\Lambda^0$ barions, and $\Omega^-$ barions (bottom panel) for Au-Au collisions at $\sqrt{s_{\rm NN}}= 200~{\rm GeV}$ and the impact parameter of $7.16$ fm.  All results were obtained  using the fluid dynamical input, obtained with event-averaged ``tilted'' initial conditions and the freeze-out temperature $T_{\rm freeze} = 150\,{\rm MeV}$. The chemical potentials $\mu_B =28.5\,{\rm MeV}$, $\mu_{I_3} =-0.9\,{\rm MeV}$, and $\mu_S =6.9\,{\rm MeV}$ were included at the freeze-out solely. The presentation of results is limited to the ``soft'' transverse momenta ($p_T<3\,{\rm GeV}$), where the fluid dynamical models are expected to be applicable, and the midrapidity region, $|y_p|<1$. One observes that slopes of the spectra are species-dependent, which is mainly due to their different masses. At intermediate momenta the spectra have exponential shapes, which is characteristic for  thermal systems. At low $p_T$ various effects play role, see next section. In the Fig.~\ref{fig:specpseu} we present the respective $p_T$-integrated pseudorapidity  distribution of charged particles, where $\eta  =   \ln\left[ (p + p_{z})/(p - p_{z})\right]/2$. The latter are compared to the contributions from $\pi^+$, $K^+$, and protons. One observes that, while the central rapidity region $y_p\approx \eta$ is approximately boost-invariant, the forward/backward rapidity regions are not. This is a result of using full four-dimensional fluid dynamical simulations of the emitting source. 
%
%
%
\subsection{Impact of resonance decays}
\label{ssec:122}
\begin{figure}[t]
\begin{center}
 \includegraphics[angle=0,width=0.65 \textwidth]{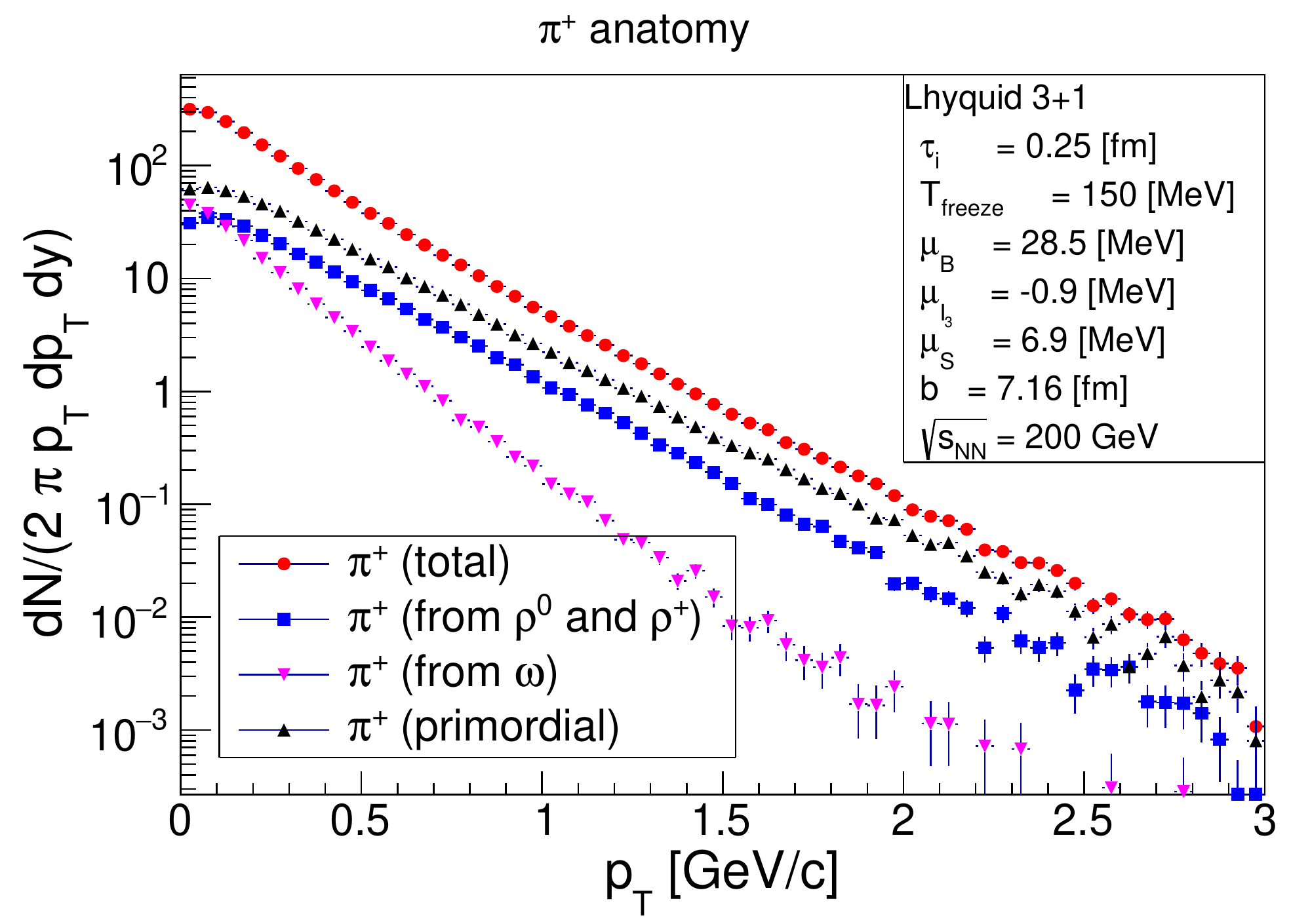}    
\end{center}
 \caption{The  anatomy of the transverse-momentum spectra of $\pi^+$ for Au-Au collisions at $\sqrt{s_{\rm NN}}= 200~{\rm GeV}$, and the impact parameter of $7.16$ fm. The spectrum of primordial pions, as well as the contribution from $\rho^0$, $\rho^+$ and $\omega$ resonance decays, is presented. The statistics of 3000 events was used. The errors are statistical only.}
\label{fig:specpion}        
\end{figure}

\sectionmark{rd}
%
One of the effects which significantly affects shapes of the single-particle spectra are decays of resonances. Due to available phase space they populate mainly the low-$p_T$ region of the spectra.
In the Fig.~\ref{fig:specpion} we present the ``anatomy'' of the transverse-momentum spectra of $\pi^+$. One observes that the low-$p_T$ part of the spectrum of primordial pions (produced directly at the freeze-out hypersurface) and the total spectrum (including the contribution from all resonance decays) differ significantly. The primordial pions develop a characteristic knee in the soft region. The decays of heavy resonances feed up the spectrum in this region, see contribution from $\omega$  decays in the  Fig.~\ref{fig:specpion}. As a result the total shape of pion spectrum takes the characteristic concave shape. Moreover, the effective slope of the final spectrum becomes steeper which manifests itself by the lower effective temperature of the spectrum, see Sec.~\ref{sec:9}.
%
%
\subsection{Experimental feed-down corrections}
\label{ssec:123}
\begin{figure}[h]
\begin{center}
 \includegraphics[angle=0,width=0.65 \textwidth]{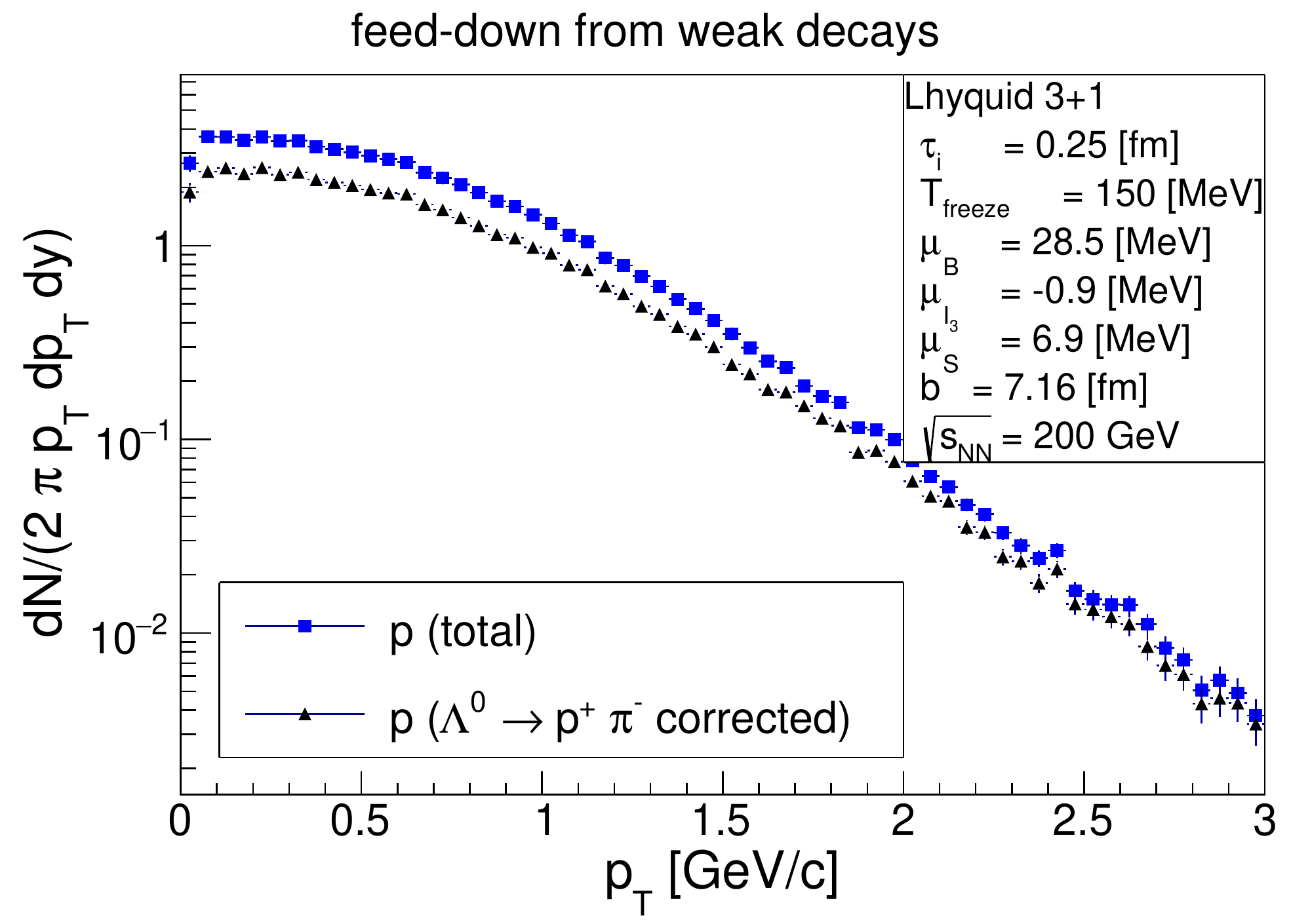}    
\end{center}
 \caption{Transverse-momentum spectra of protons, with, and without feed-down correction  for $\Lambda^0\to p^+ + \pi^-$  weak decays. The statistics of 3000 events was used. The errors are statistical only.}
\label{fig:specfeed}        
\end{figure}

\sectionmark{rd}
%
The experimental proton spectra are usually feed-down corrected for $\Lambda^0 \to p^+ + \pi^-$ weak decays. Such corrections are straightforward to be included in \THERMINATOR\, analysis, and they were also applied in Fig.~\ref{fig:spec}. In the Fig.~\ref{fig:specfeed} we present the comparison of proton spectra with and without applying these corrections. We observe that the feed-down from weak-decays is at the level of $30\%$, which is a significant correction.
%
%
%
\subsection{Ratios of particle yields}
\label{ssec:124}
\sectionmark{rd}
%
As it was discussed in Sec.~\ref{sec:9} the inclusion of  resonance decays was crucial for the proper description of the ratios of particle yields, giving the chemical freeze-out temperatures of the order of the critical temperature of the phase-transition in QCD. Following some recent studies, which focus on reproducing the shapes of the spectra rather than the yields ratios, the freeze-out temperatures extracted from the data can be as low as 150 MeV (sometimes even lower). One should expect that, in such a case, on should observe a decrease of quality of the fits of the particle abundances. To see this, in Table~\ref{table:ratios} we present the  $K^+/\pi^+$ and  $p/\pi^+$ total yield ratios calculated at various freeze-out temperatures $T_{\rm freeze}$. One observes that with reducing the freeze-out temperature the ratios decrease which is consequence of the fact that heavy particles are most copiously produced at large temperatures. The decrease is more significant for protons than for kaons. The results suggest that the fitting of particle spectra should be always accompanied by the fits of particle yields.
\begin{table}[t]
  \begin{center}
    \begin{small}
      \begin{tabular}{lcccc@{\hskip 1cm}ccc}
      \hline \\ [-1ex]
$T_{\rm freeze} \,{\rm [MeV]}$ & 130 & \multicolumn{3}{c}{150} & \multicolumn{3}{c}{170} \\  [1ex] \hline \\ 
$K^+/\pi^+$ & 0.199 & \multicolumn{3}{c}{0.263} & \multicolumn{3}{c}{0.326} \\  [2ex]
$p/\pi^+$ & 0.033 & \multicolumn{3}{c}{0.065} & \multicolumn{3}{c}{0.110} \\ \\ \hline
      \end{tabular}
    \end{small}
  \end{center}
  \caption{\small  The  $K^+/\pi^+$ and  $p/\pi^+$ total yield ratios calculated at various freeze-out temperatures $T_{\rm freeze}\in\{130, 150, 170\}\, {\rm MeV}$.
} 
  \label{table:ratios}
\end{table}

\begin{acknowledgement}
The author would like to express his gratitude to the Organizers of the \emph{53rd Karpacz Winter School of Theoretical Physics and THOR COST Action Training School} for their help and hospitality. This work was supported  by the THOR COST Action CA15213, ExtreMe Matter Institute EMMI at the GSI Helmholtzzentrum f̈ur Schwerionenforschung, Darmstadt, Germany, and the  Polish National Science Center grants No. DEC-2012/07/D/ST2/02125 and DEC-2016/23/B/ST2/00717.
\end{acknowledgement}
 
   \bibliographystyle{unsrt}
  \bibliography{references}
\end{document}